\documentclass[10pt,letterpaper]{article}
\usepackage{opex3,color,geometry}
\usepackage{amssymb}
\usepackage{amsmath}
\usepackage{graphicx}

\newcommand{\beq}{\begin{equation}}
\newcommand{\eeq}{\end{equation}}
\newcommand{\be}{\begin{enumerate}}
\newcommand{\ee}{\end{enumerate}}
\newcommand{\beas}{\begin{eqnarray*}}
\newcommand{\eeas}{\end{eqnarray*}}
\newcommand{\bea}{\begin{eqnarray}}
\newcommand{\eea}{\end{eqnarray}}

\newcommand{\vmin}{v_{\mathrm{min}}}
\newcommand{\vmax}{v_{\mathrm{max}}}

\newcommand{\rmd}{\mathrm{d}}

\newcommand{\gc}{\gamma_c}
\newcommand{\gcp}{\gamma'_c}

\begin{document}

\title{
Full vectorial analysis of polarization effects in optical nanowires 
}

\author{Shahraam Afshar V.*, M. A. Lohe, Wen Qi Zhang, and
\\
Tanya M. Monro}
\address{Institute for Photonics \& Advanced Sensing (IPAS),
The University of Adelaide, 5005, Australia}

\email{*shahraam.afshar@adelaide.edu.au}


\begin{abstract}
We develop a full theoretical analysis of the nonlinear interactions 
of the two polarizations of a waveguide by means of a vectorial
model of pulse propagation which applies to 
high index subwavelength wave\-guides. In
such waveguides there is an anisotropy in the nonlinear behavior of the two
polarizations that originates entirely from the waveguide structure,
and leads to switching properties. We determine the stability
properties of the steady state solutions by means of a Lagrangian formulation. 
We find all static solutions of the nonlinear system, including
those that are periodic with respect to the optical fiber length
as well as nonperiodic soliton solutions, and analyze these solutions
by means of a Hamiltonian formulation. 
We discuss in particular the switching solutions which lie near
the unstable steady states, since they lead to 
self-polarization flipping which can in principle be employed to construct 
fast optical switches and optical logic gates.
\end{abstract}





\section{Introduction}

The Kerr nonlinear interaction of the two polarizations of the
propagating modes of a  wave\-guide leads to a host of physical effects that
are significant from both fundamental and application points of view.
Here, we develop a model of nonlinear interactions of the two 
polarizations using full vectorial nonlinear pulse propagation equations, 
with which we analyze the nonlinear interactions in the 
emerging class of subwavelength and high index 
optical  wave\-guides. Based on this model we predict an anisotropy 
that originates solely from the  wave\-guide structure, and which
leads to switching states that can in principal be used to construct 
optical devices such as switches or logical gates.
We derive the underlying nonlinear Schr\"{o}dinger equations of the 
vectorial model with explicit integral expressions for the nonlinear
coefficients. We analyze solutions of these nonlinear pulse propagation 
equations and the associated switching states by means of a Lagrangian
formulation, which enables us to determine stability properties
of the steady states; this formulation provides a global view of all
solutions and their properties by means of the potential function and leads, 
for example, to the emergence of kink solitons as solutions to the model 
equations. We also use a Hamiltonian formalism in order to identify periodic 
and solitonic trajectories, including solutions that allow polarization 
flipping, and find conditions under which the unstable states and
associated switching solutions are experimentally accessible.

The nonlinear interactions of the two polarizations
of the propagating modes of a  wave\-guide
have been studied extensively over the last 30 years
\cite{Stolen82}-\cite{Kozlov11}.
Different aspects of the interactions have been investigated, for example
Stolen et al.\  \cite{Stolen82}
used the induced nonlinear phase difference between the two polarizations to
discriminate between high and low power pulses. In the context of
counterpropagating waves, the nonlinear interactions
have been shown to lead to polarization domain wall solitons, 
\cite{Pitois98}-\cite{Wabnitz09}, \cite{Zakharov87}
which are
described as kink solitons representing a polarization switching between
different domains with orthogonal polarization states.
The nonlinear interactions can also lead to polarization attraction
\cite{Pitois01}, \cite{Kozlov10}-\cite{Kozlov11}, \cite{Pitois05,Pitois08}
where the state
of the polarization of a signal is attracted towards that of a pump beam. 
For twisted birefringent optical fibers, polarization instability
\cite{Matera86,Feldman93} and polarization domain wall solitons
\cite{Wabnitz09b} have been reported. 
The nonlinear interactions also induce modulation
instability which results in dark-soliton-like pulse-train generation
\cite{Millot97,Millot98}. Large-signal enhanced frequency conversion
\cite{Seve98}, cross-polarization modulation for WDM signals \cite{Wabnitz09},
and polarization instability \cite{Winful86} have also been reported and
attributed to nonlinear polarization interactions.
Stability behavior has been studied in
anisotropic crystals \cite{Gregori86}.

The nonlinear interactions of the two polarizations can also be studied in the
context of either 
\emph{nonlinear coherent coupling} or 
\emph{nonlinear directional coupling} 
in which the amplitudes of two or more electric fields, either
the two polarizations of a propagating mode of a  wave\-guide or different 
modes
of different  wave\-guides, couple to each other through linear and nonlinear
effects 
\cite{Jensen82}-\cite{Wang06}. 
Nonlinear directional coupling is relevant to ultrafast all-optical 
switching, such as soliton switching
\cite{Kivshar93}-\cite{Khan08}
and all-optical logic gates 
\cite{Yang91}-\cite{Fraga06}. 
The interaction of ultrafast
beams, with different frequencies and polarizations, in anisotropic media has
also been studied and the conditions for polarization stability have been
identified \cite{Hutchings97, Hutchings95}.

In previous work (\cite{Agrawal07}, Chapter 6), 
the nonlinear interactions of the two polarizations are
described by two coupled Schr\"{o}dinger equations. These equations employ
the weak guidance approximation, which assumes that the
propagating modes of the two polarizations of the  wave\-guide are purely
transverse and orthogonal to each other within the transverse $x,y$ plane, 
perpendicular to the direction of propagation $z$. Based on this, the
electric fields are written as 
\beq
\mathbf{E}_{i}(x,y,z,t)=A_{i}(z,t)\mathbf{e}_{i}(x,y), \quad
i=1,2, 
\eeq
where $A_i(z,t)$ are the amplitudes of the two polarizations, with
$\mathbf{e}_{1}(x,y)\centerdot\mathbf{e}_{2}(x,y)=
e_{1}(x,y)e_{2}(x,y)\mathbf{\hat{x}}\centerdot \mathbf{\hat{y}}=0$, 
where $e_1(x,y),e_2(x,y)$
are the transverse distributions of the two polarizations, 
$\mathbf{\hat{x},\hat{y}}$ are unit vectors along the $x$ and $y$ 
directions, and it is understood that fast oscillatory terms of the form
$\exp(-i\omega t\pm\beta_{i}z)$ are to be included for the polarization fields.
The weak guidance approximation also assumes that the Kerr nonlinear
coefficients for the self phase modulation of the two polarizations are equal
because their corresponding mode effective areas are equal
\cite{Agrawal07}. We refer here to models of nonlinear pulse propagation
based on the weak guidance approximation simply as ``scalar" models, 
since these models consider only purely transverse modes for the two
polarizations.

The weak guidance approximation works well only for  wave\-guides with low 
index
contrast materials, and large dimension structure compared to the operating
wavelength. This approximation is, however, no longer appropriate for high
index contrast subwavelength scale wave\-guides (HIS-WGs) 
\cite{Afshar09}-\cite{Daniel10}. 
These
wave\-guides have recently attracted significant interest mainly due to their
extreme nonlinearity and possible applications for all optical photonic-chip
devices. Examples include silicon, chalcogenide, or soft
glass optical  wave\-guides, which have formed the base for three active field
 of studies: silicon photonics 
\cite{Almeida04}-\cite{Lau11},
chalcogenide photonics
\cite{Pelusi09}-\cite{Eggleton11}, 
and soft glass microstructured photonic devices
\cite{Petropoulos01}-\cite{Poletti11}.

In order to address the limitations of the scalar models 
in describing nonlinear processes in HIS-WGs,
we have developed in \cite{Afshar09} a full vectorial nonlinear pulse 
propagation model. 
Important features of this model are: (1) the propagating modes of the
 wave\-guide are not, in general, transverse and have large $z$ components and, 
(2) the orthogonality condition of different
polarizations over the cross section of the  wave\-guide is given by
$\int\mathbf{e}_{1}(x,y)\times\mathbf{h}_{2}^{\ast}(x,y)\centerdot
\mathbf{\hat{z}}\;\rmd A=0$,
rather than simply $\mathbf{e}_{1}(x,y)\centerdot\mathbf{e}_{2}(x,y)=0$ as
in the scalar models. These aspects lead to an improved understanding of many
nonlinear effects in HIS-WGs; it was predicted in \cite{Afshar09}, 
for example, that within the vectorial model 
the Kerr effective nonlinear coefficients of HIS-WGs
have higher values than those predicted by the scalar models 
due to the contribution of
the $z$-component of the electric field, as later confirmed
experimentally \cite{Afshar09c}. Similarly, it was also predicted that
modal Raman gain of HIS-WGs should be higher than expected from 
the scalar model \cite{Turner09}.

Here, we extend the vectorial model to investigate the nonlinear
interaction of the two polarizations of a guided mode. 
The full vectorial model leads to an induced anisotropy on the dynamics of 
the nonlinear interaction of the two polarizations \cite{Zhang:11}, 
which we refer to as
structurally induced anisotropy, in order to differentiate
this anisotropy from others, such as those for which the anisotropy 
originates from isotropic materials. 
The origin of the anisotropy is the structure of the  wave\-guide
rather than the  wave\-guide material. 

The origin of this anisotropy in subwavelength and high index contrast 
waveguides has also been reported by Daniel and Agrawal \cite{Daniel10},
who considered nonlinear interactions of the two polarizations 
in a silicon rectangular nanowire including the effect of free carriers. 
In their analysis, however, they ignore the coherent coupling of the two 
polarizations, considering the dynamics  of the Stokes parameters only for a 
specific waveguide and ignore the linear phase.  

This anisotropy in turn leads to a new parameter space
in which the interaction of the two polarizations shows switching behavior,
which is a feature of the vectorial model not accessible through the 
scalar model with the underlying weak guidance approximation.
We also show that the resulting system of nonlinear equations,
for the static case, can be solved analytically. Due to the
underlying similarity of the nonlinear interaction of the 
two polarizations and the
nonlinear directional coupling of two  wave\-guides, the anisotropy discussed
here can be also applied to the case of nonlinear directional coupling,
in which the two wave\-guides have different
effective nonlinear coefficients for the propagating modes.

This work develops and expands on results reported for the first
time in \cite{Zhang:11,Afshar09d}, in particular we derive here
(in Section \ref{theory}) the equations that describe the nonlinear
interactions of the two polarizations within the framework of the
vectorial model, with explicit integral expressions for the nonlinear
coefficients. In Section \ref{ss1} we determine properties of the
static solutions, classify the steady state solutions, and determine 
their stability using a Lagrangian formalism. 
We also discuss a Hamiltonian approach and how the phase space portrait
provides a complete picture of the trajectories of the system,
including the periodic and solitonic solutions (Section \ref{ss3}). 
We derive analytical periodic solutions by direct integration of the system
of equations in Section \ref{ss4}, and then discuss switching solutions and
their properties. We relegate to the Appendix a mathematical analysis of the
exact soliton solutions, which are relevant to the switching solutions,
with concluding remarks in Section \ref{s9}. 


\section{Nonlinear differential equations of the model\label{theory}}

In the vectorial model the nonlinear pulse propagation of different modes of a
 wave\-guide is described by the equations:
\bea
\nonumber
\frac{\partial A_{\nu}}{\partial z}
&+&
\sum_{n=1}^{\infty}\frac{i^{n-1}\beta_{\nu}^{(n)}}{n!}
\frac{\partial^{n}A_{\nu}}{\partial t^{n}}
\\
\nonumber
&=&
i\left(\gamma_{\nu}\left|A_{\nu}\right|^2
+\gamma_{\mu\nu}\left|A_{\mu}\right|^2\right)A_{\nu}
+i\gamma'_{\mu\nu}A_{\mu}^2A_{\nu}^{\ast}e^{-2i(\beta_{\nu}-\beta_{\mu})z}
+i\gamma_{\mu\nu}^{(1)}A_{\mu}^{\ast}A_{\nu}^2
e^{-i(\beta_{\mu}-\beta_{\nu})z}
\\
&&
+i\gamma_{\mu\nu}^{(2)}A_{\mu}\left|A_{\nu}\right|^2
e^{i(\beta_{\mu}-\beta_{\nu})z}
+i\gamma_{\mu\nu}^{(3)}A_{\mu}\left|A_{\mu}\right|^2
e^{i(\beta_{\mu}-\beta_{\nu})z}
\label{01}
\eea 
where $\mu,\nu=1,2$ with $\mu\ne\nu$, and
$A_1(z,t),A_2(z,t)$ are the amplitudes of the two orthogonal polarizations.
These equations follow from the analysis in \cite{Afshar09}, by
combining  Eqs.\ (23,32) of \cite{Afshar09}, but without the shock term.
The linear birefringence is defined by
 $\Delta\beta_{\nu\mu}=-\Delta \beta_{\mu\nu}=\beta_{\nu}-\beta_{\mu}$ 
and the $\gamma$ coefficients are given by
\bea
\gamma_{\nu}
&=&
\left(\frac{k\varepsilon_0}{4\mu_0}\right)
\frac{1}{3N_{\nu}^2}\int n^2(x,y)n_2(x,y)
\left[2\left|\mathbf{e}_{\nu}\right|^4
+
\left|\mathbf{e}_{\nu}^2\right|^2
\right]\rmd A,
\label{100}
\\
\gamma_{\mu\nu}  
&=&
\left(\frac{k\varepsilon_0}{4\mu_0}\right)
\frac{2}{3N_{\nu}N_{\mu}}\int n^2(x,y)n_2(x,y)
\left[\left\vert\mathbf{e}_{\nu}\centerdot\mathbf{e}_{\mu}^{\ast}\right\vert^2
+\left\vert \mathbf{e}_{\nu}\centerdot\mathbf{e}_{\mu}\right\vert^2
+\left\vert \mathbf{e}_{\nu}\right\vert^2
\left\vert \mathbf{e}_{\mu}\right\vert^2
\right]\rmd A,
\label{100a}
\\
\gamma'_{\mu\nu} 
&=&
\left(\frac{k\varepsilon_0}{4\mu_0}\right)
\frac{1}{3N_{\nu}N_{\mu}}\int n^2(x,y)n_2(x,y)
\left[2(\mathbf{e}_{\mu}\centerdot\mathbf{e}_{\nu}^{\ast})^2
+(\mathbf{e}_{\mu})^2(\mathbf{e}_{\nu})^2
\right]\rmd A,
\label{100b}
\\
\gamma_{\mu\nu}^{(1)} 
&=&
\left(\frac{k\varepsilon_0}{4\mu_0}\right)
\frac{1}{3\sqrt{N_{\nu}^{3}N_{\mu}}}\int n^2(x,y)n_2(x,y)
\left[
2\left\vert \mathbf{e}_{\nu}\right\vert^2(\mathbf{e}_{\mu}^{\ast}\centerdot
\mathbf{e}_{\nu})
+(\mathbf{e}_{\nu})^2(\mathbf{e}_{\mu}^{\ast}\centerdot
\mathbf{e}_{\nu}^{\ast})
\right]\rmd A,
\label{100c}
\\
\gamma_{\mu\nu}^{(2)} 
&=&
\left(\frac{k\varepsilon_0}{4\mu_0}\right)
\frac{2}{3\sqrt{N_{\nu}^{3}N_{\mu}}}\int n^2(x,y)n_2(x,y)
\left[
2\left\vert \mathbf{e}_{\nu}\right\vert^2
(\mathbf{e}_{\mu}\centerdot\mathbf{e}_{\nu}^{\ast})
+(\mathbf{e}_{\nu}^{\ast})^2(\mathbf{e}_{\mu}\centerdot\mathbf{e}_{\nu})
\right]\rmd A,
\label{100d}
\\
\gamma_{\mu\nu}^{(3)}  
&=&
\left(\frac{k\varepsilon_0}{4\mu_0}\right)
\frac{1}{3\sqrt{N_{\mu}^{3}N_{\nu}}}\int n^2(x,y)n_2(x,y)
\left[
2\left\vert\mathbf{e}_{\mu}\right\vert^2
(\mathbf{e}_{\mu}\centerdot\mathbf{e}_{\nu}^{\ast})+
(\mathbf{e}_{\mu})^2(\mathbf{e}_{\mu}^{\ast}\centerdot\mathbf{e}_{\nu}^{\ast})
\right]\rmd A.
\label{100e}
\eea
Here we use the notation 
$(\mathbf{e}_{\nu})^2=\mathbf{e}_{\nu}\centerdot\mathbf{e}_{\nu},
|\mathbf{e}_{\nu}|^2=\mathbf{e}_{\nu}\centerdot\mathbf{e}_{\nu}^{\ast}$
and
$\left|\mathbf{e}_{\nu}^2\right|^2=(\mathbf{e}_{\nu}\centerdot\mathbf{e}_{\nu})
(\mathbf{e}_{\nu}^{\ast}\centerdot\mathbf{e}_{\nu}^{\ast})$,
together with
$\left\vert \mathbf{e}_{\nu}\centerdot\mathbf{e}_{\mu}^{\ast}%
\right\vert^{2} =(\mathbf{e}_{\nu}.\mathbf{e}_{\mu}^{\ast})%
(\mathbf{e}_{\nu}^{\ast}\centerdot\mathbf{e}_{\mu})$.
In these equations $\mathbf{e}_1(x,y), \mathbf{e}_2(x,y)$ 
are the modal fields of the two orthogonal polarizations,
$k=2\pi/\lambda$ is the propagation constant in vacuum, and $\gamma_{\nu}$,
$\gamma_{\mu\nu}$, $\gamma_{\mu\nu}^{\prime},$ $\gamma_{\mu\nu}^{(1)},$
$\gamma_{\mu\nu}^{(2)},\gamma_{\mu\nu}^{(3)}$ are the effective
nonlinear coefficients representing, respectively, 
self phase modulation, cross phase modulation,  
and coherent coupling of the two polarizations, and
\beq
N_{\mu}=\frac{1}{2}
\left\vert\int\mathbf{e}_{\mu}\times\mathbf{h}_{\mu}^{\ast}
\centerdot\widehat{\mathbf{z}}\;\rmd A\right\vert 
\eeq
is the normalization parameter. 

The
coupled equations (\ref{01}) describe the full vectorial nonlinear interaction
of the two polarizations. There are two fundamental differences between these
equations and the typical scalar coupled Schr\"odinger equations 
(see for example Chapter 6 in \cite{Agrawal07}). 
Firstly, the additional terms 
$A_{\mu}^{\ast}A_{\nu}^2, A_{\mu}\left\vert A_{\nu}\right\vert^2,
A_{\mu}\left\vert A_{\mu}\right\vert^2$
on the right hand side of Eq.\ (\ref{01})
represent interactions between the two polarizations. These do not 
appear in the scalar model since the effective nonlinear 
coefficients associated with these terms,
$\gamma_{\mu\nu}^{(1)},\gamma_{\mu\nu}^{(2)},\gamma_{\mu\nu}^{(3)}$ 
as given in Eqs.\ (\ref{100c},\ref{100d},\ref{100e}),
contain factors such as
$\mathbf{e}_{\mu}\centerdot\mathbf{e}_{\nu}$ which
are zero in the scalar model, since the modes are assumed to be purely
transverse. 
All possible third power combinations of the two polarization fields, namely
$\left\vert A_{\nu}\right\vert^2A_{\nu},
\left\vert A_{\mu}\right\vert^2A_{\nu},
A_{\mu}^2A_{\nu}^{\ast},A_{\mu}^{\ast}A_{\nu}^2,
A_{\mu}\left\vert A_{\nu}\right\vert^2$ and 
$A_{\mu}\left\vert A_{\mu}\right\vert^2$ 
occur on the right hand side of Eq.\ (\ref{01}), 
due to the $z$-component of the modal fields.
Secondly, in all effective nonlinear coefficients given by 
Eqs.\ (\ref{100}-\ref{100e}),
the modal fields $\mathbf{e}$ and $\mathbf{h}$ have both transverse and
longitudinal components, unlike the scalar model
in which modal fields have only transverse components. The
terms containing
nonzero $\mathbf{e}_{\mu}\centerdot\mathbf{e}_{\nu}$ 
provide a mechanism for the
interaction of the two polarizations since they allow for exchange of power
between the two modes through the $z$-components of their fields. 
The last term on the right hand side of Eq. (\ref{01}), for example,  
indicates a coupling of power into a polarization, even if initially no 
power is coupled into that polarization.

Although the terms on the right hand side of Eq.\ (\ref{01}) that contain 
$\mathbf{e}_{\mu}\centerdot\mathbf{e}_{\nu}$ are nonzero, they are 
generally significantly smaller than the remaining terms and are 
therefore neglected in the following; further
investigation of the effects of these terms, and a discussion of their
physical significance, will
be presented elsewhere. The focus of this paper is to investigate the
effect of the $z$-components of the fields $\mathbf{e}$ and $\mathbf{h}$, 
which influence the values of the effective coefficients, and therefore also 
the nonlinear interactions of the two polarizations. Hence, from (\ref{01}),
we obtain the equations:
\begin{equation}
\frac{\partial A_{\nu}}{\partial z}
+
\sum_{n=1}^{\infty}\frac{i^{n-1}}{n!}
\beta_{\nu}^{(n)}
\frac{\partial^{n}A_{\nu}}{\partial t^{n}}
=
i\left(\gamma_{\nu}\left\vert A_{\nu}\right\vert ^2
+\gamma_{\mu\nu}\left\vert A_{\mu}\right\vert ^2\right)A_{\nu}
+i\gamma'_{\mu\nu}A_{\mu}^2A_{\nu}^{\ast}
e^{-2i(\beta_{\nu}-\beta_{\mu})z}. 
\label{04}
\end{equation}
These are similar in form to the scalar coupled equations 
(\cite{Agrawal07}, Section 6.1.2), however, the
coefficients $\gamma_{\nu},\gamma_{\mu\nu},\gamma_{\mu\nu}^{^{\prime}}$,
given in Eqs.\ (\ref{100}-\ref{100b}), now contain $z$-components of the
electric field, through both $\mathbf{e}$ and $\mathbf{h}$. 
In the framework of the scalar model, the weak guidance approximation 
assumes that the effective mode areas of the two polarization modes are 
equal \cite{Agrawal07}, leading to
\begin{equation}
\gamma_{1} = \gamma_{2} = 3\gamma_{c}/2 = 3\gamma_{c}^{\prime},
\label{eq02}
\end{equation}
where we have denoted
$\gc=\gamma_{12}=\gamma_{21},\gcp=\gamma'_{12}=\gamma'_{21}$.
This means that in the scalar model there is an isotropy
of the nonlinear interaction of the two polarizations; in order
to break this isotropy,
one needs to use either anisotropic  wave\-guide materials or twisted
fibers, or else
couple varying light powers into the two polarizations by
using either counter- or co-propagating laser beams. 
The fact that in the vectorial form (\ref{04}) of the coupled equations 
the $\gamma$
values include the $z$-component of the fields, as given by
Eqs.\ (\ref{100}-\ref{100b}), means that the equalities (\ref{eq02}) do not
hold in general.  As an example, see Fig.\  1 in \cite{Zhang:11} which
plots $\gamma_1,\gamma_2,\gc,\gcp$ for a 
step-index glass-air  wave\-guide with an elliptical cross
section; evidently the equalities (\ref{eq02}) are not satisfied.
One consequence of the vectorial formulation is, 
as we show in Section \ref{stab}, the existence of unstable states not 
present in the scalar  formulation. 


\section{Static equations\label{ss1}}

We find now all solutions of Eq.\ (\ref{04}) for the static case, in which
the fields $A_1,A_2$ are functions of $z$ only. We have therefore
the two equations
\bea
\label{f1}
\frac{d A_1}{d z}
&=&
i\left(\gamma_1\left\vert A_1\right\vert^2
+\gc\left\vert A_2\right\vert ^2\right)A_1
+i\gcp A_2^2A_1^{\ast}\,e^{-2i\Delta\beta\, z}
\\
\label{f2}
\frac{d A_2}{d z}
&=&
i\left(\gamma_2\left\vert A_2\right\vert^2
+\gc\left\vert A_1\right\vert^2\right)A_2
+i\gcp A_1^2A_2^{\ast}\, e^{2i\Delta\beta\, z},
\eea
where $\Delta\beta=\beta_1-\beta_2$.
We express
the fields $A_1,A_2$ in polar form according to
\beq
\label{e30}
A_1  =\sqrt{P_1}\,e^{i\phi_1},
\qquad
A_2 =\sqrt{P_2}\,e^{i\phi_2},
\eeq
where the powers $P_1,P_2$ and the phases $\phi_1,\phi_2$ are real functions
of $z$. It is convenient to define the phase difference $\Delta\phi$ and
an angle $\theta$ according to
\beq
\Delta\phi=\phi_1-\phi_2+z\Delta\beta,\qquad
\theta=2\Delta\phi,
\eeq
then upon substitution into Eqs.\ (\ref{f1},\ref{f2}) we obtain
the four real equations:
\bea
\label{f3}
\frac{d P_1}{d z}
&=&
2\gcp P_1P_2\sin\theta
\\
\label{f4}
\frac{d P_2}{d z}
&=&
-2\gcp P_1P_2\sin\theta
\\
\label{f5}
\frac{d \theta}{d z}
&=&
2\Delta\beta
+2P_1(\gamma_1-\gc-\gcp\cos\theta)
-2P_2(\gamma_2-\gc-\gcp\cos\theta)
\\
\label{f6}
\frac{d \phi_1}{d z}
&=&
\gamma_1P_1+P_2(\gc+\gcp\cos\theta).
\eea
The last equation decouples from the remaining equations, hence
we first solve Eqs.\ (\ref{f3}-\ref{f5}) for $P_1,P_2,\theta$ 
and then determine $\phi_1$ by integrating (\ref{f6}). 
Eqs.\ (\ref{f3},\ref{f4}) show that $P_0=P_1+P_2$ is constant in $z$.
We define the dimensionless variables
\beq
\label{p1}
v=\frac{P_1}{P_0}=\frac{P_1}{P_1+P_2},
\qquad
\tau=2\gcp P_0\,z,
\eeq
and the dimensionless parameters
\beq
\label{p2}
a=-\frac{\Delta\beta}{\gcp P_0}-\frac{\gc-\gamma_2}{\gcp},
\qquad
b=\frac{\gamma_1+\gamma_2-2\gc}{2\gcp}.
\eeq

In terms of these parameters we obtain the two equations:
\begin{align}
\dot{v} & \equiv\frac{dv}{d\tau}=v(1-v)\sin\theta,
\label{e4}
\\
\dot{\theta} & \equiv\frac{d\theta}{d\tau}=-a+2bv+(1-2v)\cos\theta.
\label{e5}
\end{align}
Since $\tau$ takes only positive values, we may regard 
$\tau$ as a time variable which is limited in value only
by the length of the optical fiber and by the value of $P_0$, and we set the
initial values $v_0=v(0), \theta_0=\theta(0)$ at time $\tau=0$, i.e.\ at one 
end of the fiber. 
The general solution depends on the initial values $v_0, \theta_0$ and
on only two parameters $a,b$, even
though Eqs.\ (\ref{f3}-\ref{f6}) depend on the five constants 
$P_0, \gamma_1,\gamma_2,\gc,\gcp$.

At the initial time we have $P_1,P_2>0$ and so we always choose
$v_0$ such that $0<v_0<1$. It may be shown from Eqs.\ (\ref{e4}-\ref{e5})
that $0<v(\tau)<1$ is then maintained for all $\tau>0$, i.e.\ the powers
$P_1,P_2$ remain strictly positive at all later times. 
The constraint $0<v_0<1$ implies that the initial speed $\dot\theta_0$ 
is restricted, since it follows from Eq.\ (\ref{e5}) that 
$|\dot\theta|\leqslant |a|+2|b|+1$ at all times $\tau$.


\subsection{Properties of $a,b$\label{sss2}}

Of the two dimensionless parameters $a,b$, 
evidently $b$ depends only on the optical fiber parameters, whereas
$a$ depends also on the total power $P_0$, unless $\Delta\beta=0$.
For the scalar model, when Eqs.\ (\ref{eq02}) are satisfied, we have 
$b=1$ but generally $b\ne1$. In this case a set of
steady state solutions appears (the states (\ref{a}) discussed in Section 
\ref{sss1} below) which for certain values of $a,b$ are unstable.
For fibers with elliptical cross sections we find that
$b>1$ and the unstable steady states exist
provided $1<a<2b-1$. We have not, however,  been able to eliminate the
possibility that $b<1$ for other geometries, and so in the following 
we also analyze the case $b<1$.  
The parameter $a$ can be positive or negative
depending on the sign of $\Delta\beta$ and on the value $P_0$;
when Eqs.\ (\ref{eq02}) are satisfied we have
$a=-3\Delta\beta/(P_0\gamma_1)+1$ and hence $a$ can
take large positive or negative values for small $P_0$.

\begin{figure}[htbp]
\centering
\includegraphics[width=0.49\columnwidth]{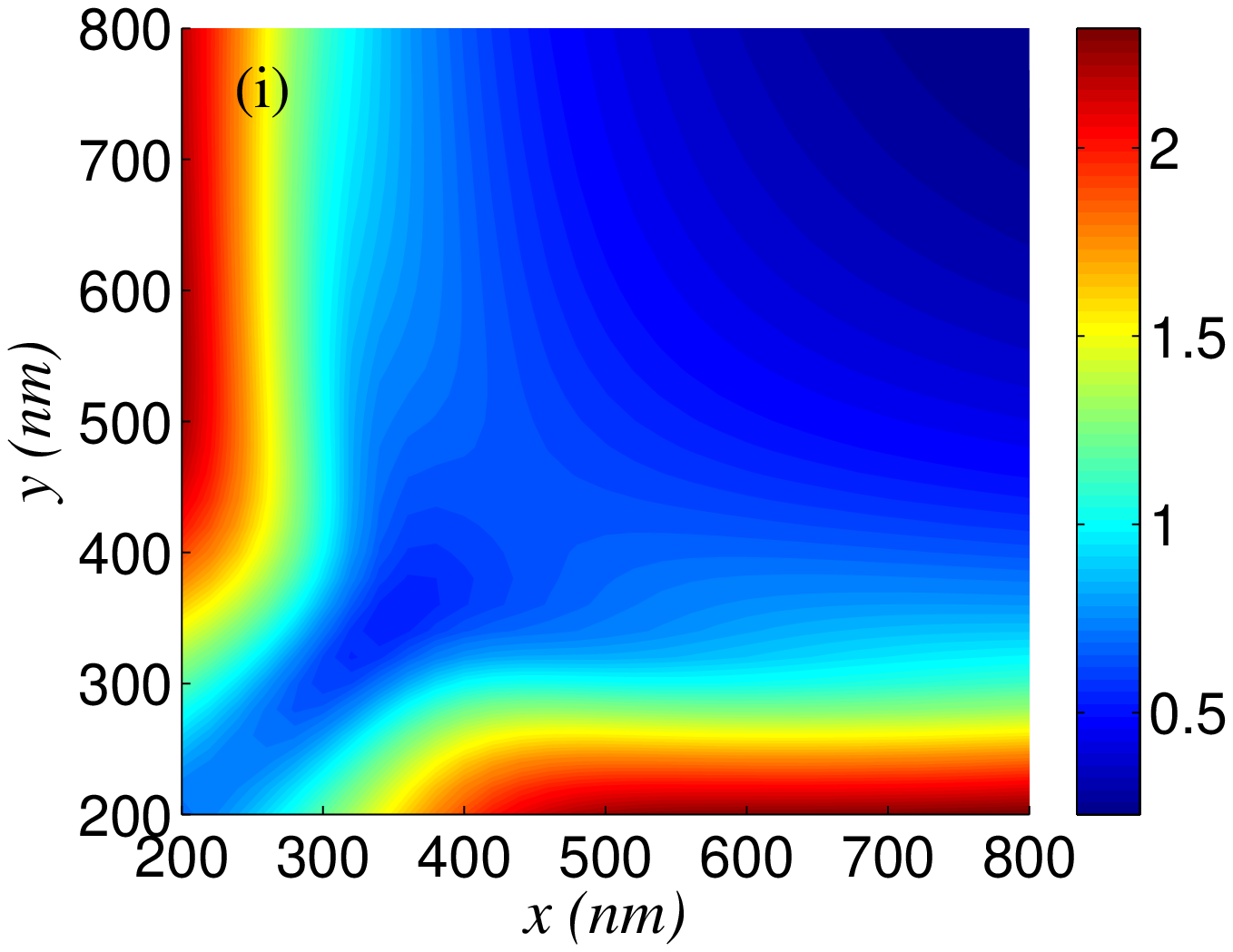}
\includegraphics[width=0.49\columnwidth]{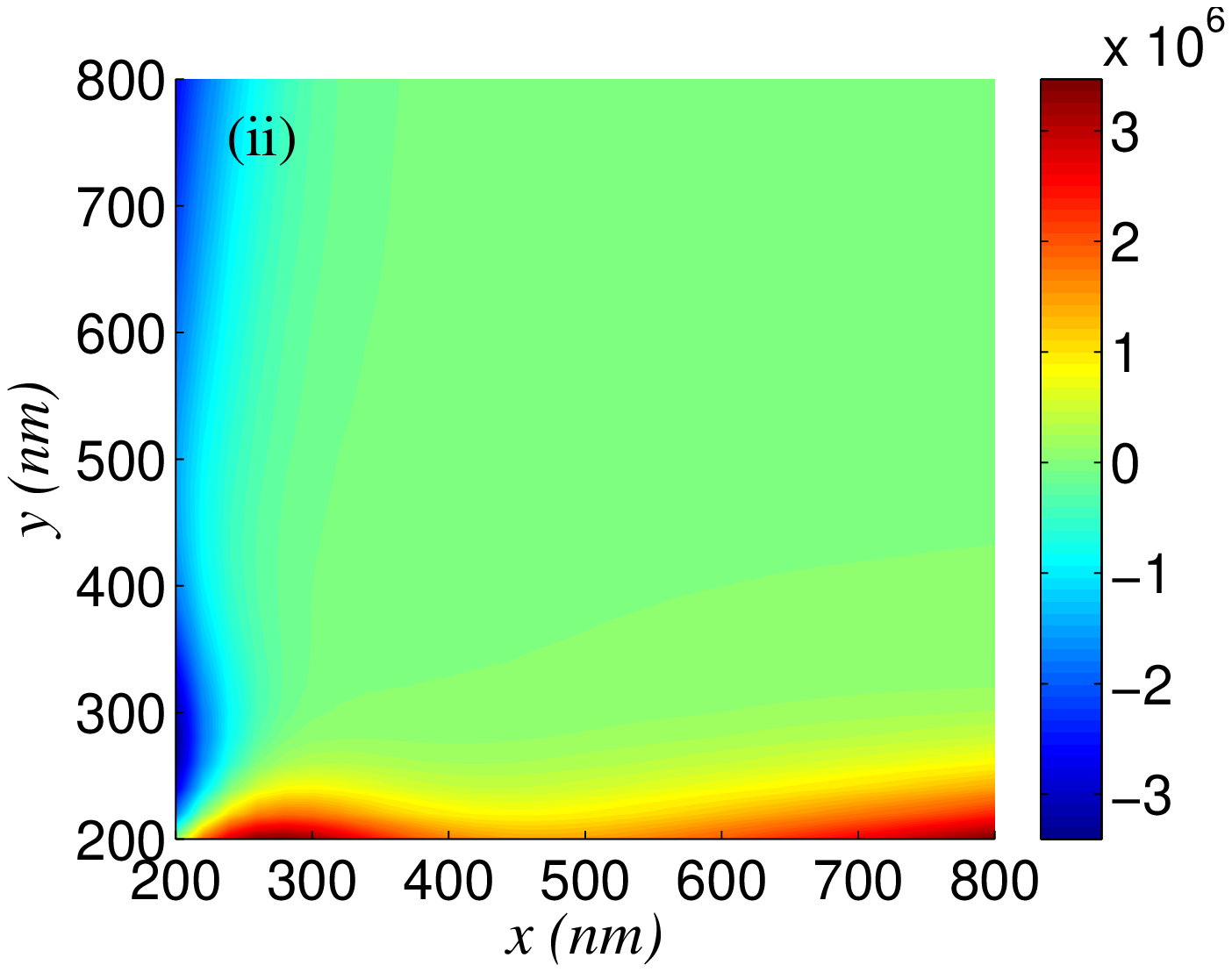}
\includegraphics[width=0.49\columnwidth]{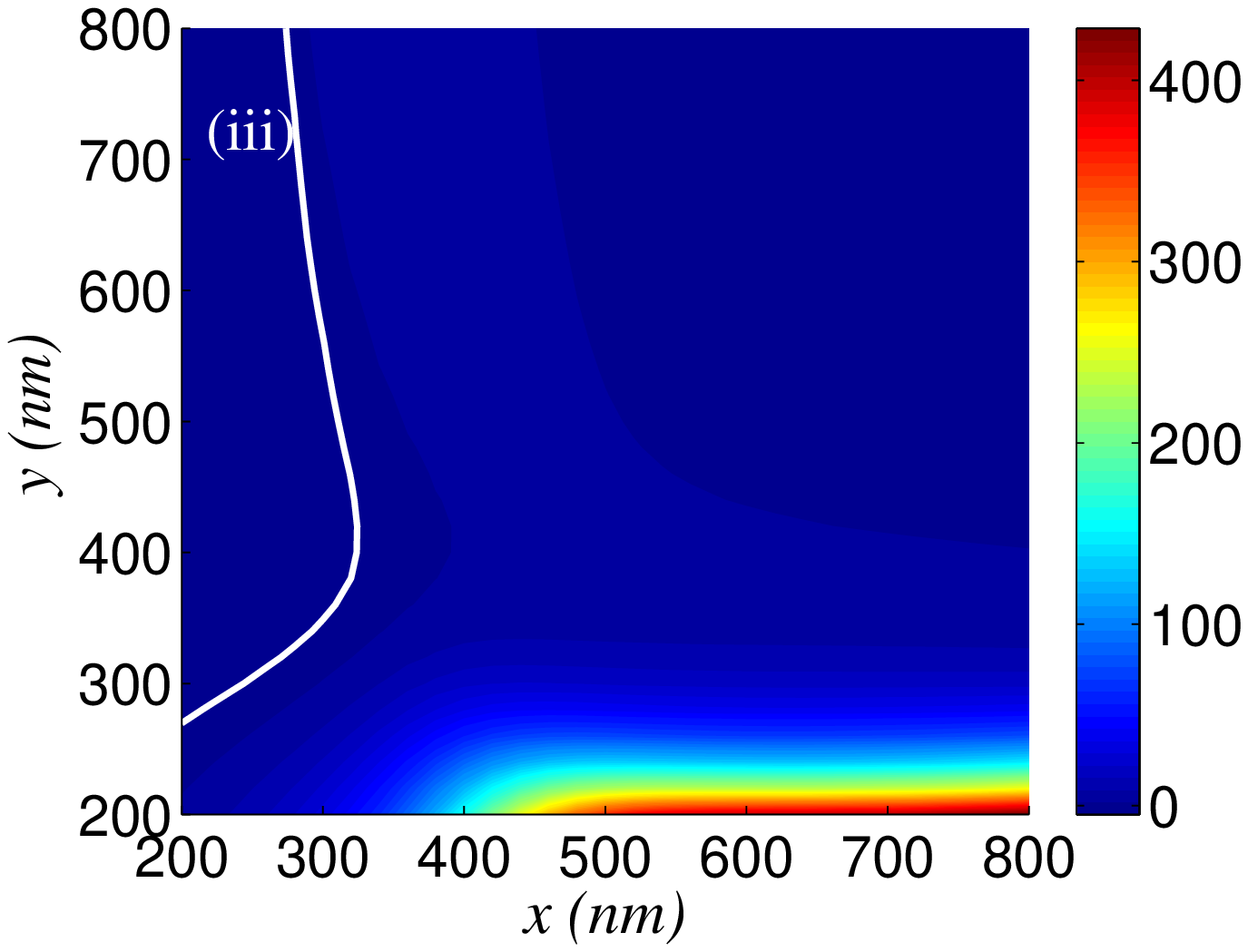}
\includegraphics[width=0.49\columnwidth]{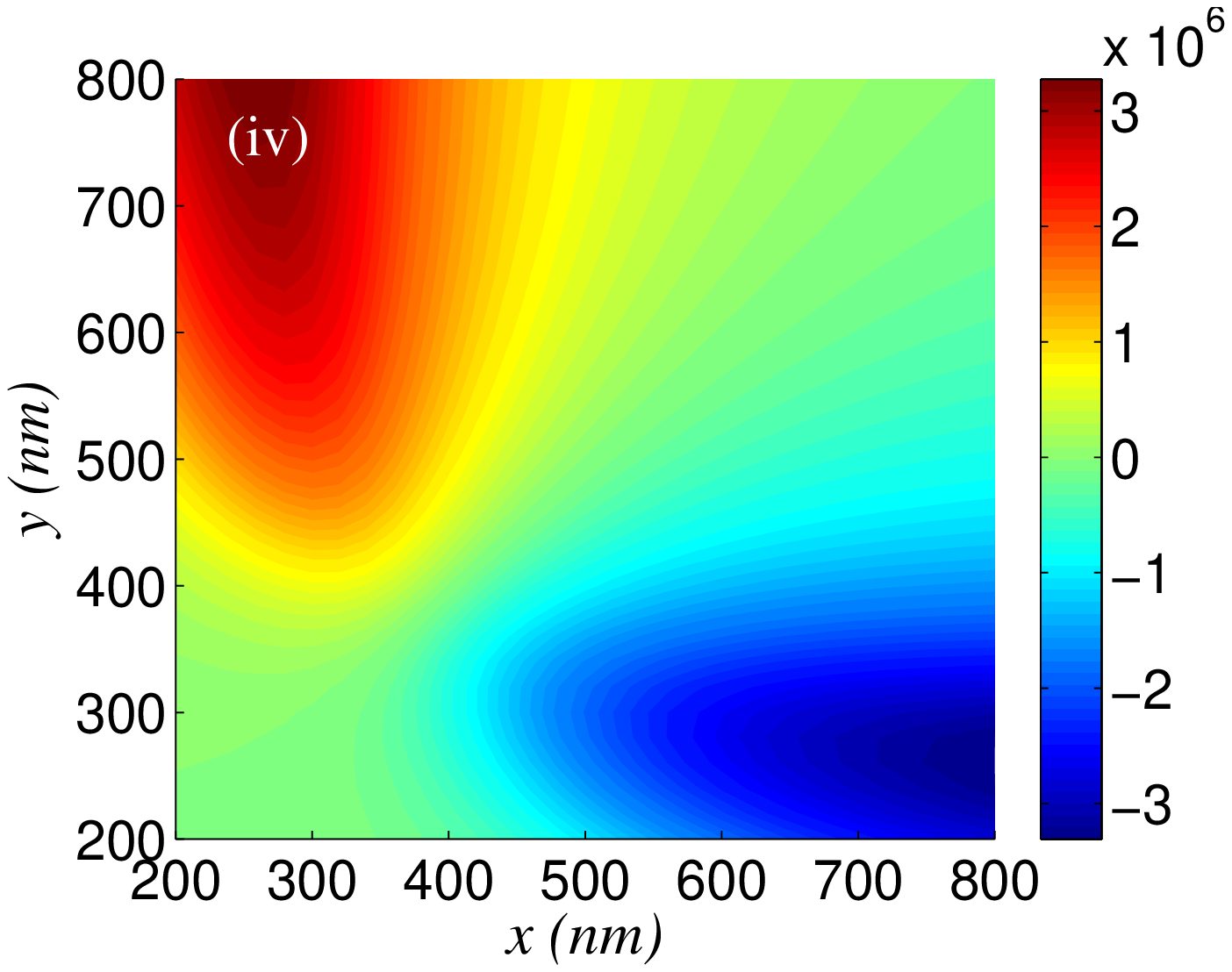}
\caption{
Contour plots as functions of the elliptical waveguide
dimensions $x,y$ of (i) $\log_{10}b$; (ii) $a$ as defined in (\ref{p2})
for $P_0=1$W;
(iii) $C=(\gamma_2-\gc)/\gcp$ where $C<0$ to the left of the white line; 
(iv) the birefringence $\Delta\beta$.
}
\label{fig1}
\end{figure}

As an example, we have evaluated $b$ using the definitions
Eqs.\ (\ref{100}-\ref{100b}) for step-index, air-clad glass
waveguides with elliptical cross sections where the major/minor axes are
denoted $x,y$.  The host glass is taken to be chalcogenide with linear and 
nonlinear refractive indices of 
$n = 2.8$ and $n_{2} = 1.1 ×\times 10^{-17}$m$^{2}$/W 
at $\lambda = 1.55\mu$m (as in \cite{MF2007}).
Fig.\ \ref{fig1}(i) shows a contour plot of $\log_{10} b$ as a 
function of $x,y$. We see, as expected, that $b$ approaches
$1$ as the waveguide dimensions $x,y$ increase towards the operating
wavelength. For small core wave\-guides, however, we find $b>1$ with 
values as large as $b\approx200$. 
The parameter $a$, on the other hand, depends on
both the structure and the total input power $P_0$. For low input powers,
specifically for
$P_0\gcp\ll|\Delta\beta|$, $a$ can take large negative
values (for
$\Delta\beta>0$) or positive values (for $\Delta\beta<0$) as shown in Fig.\
\ref{fig1}(ii). For large values of $P_0$, however, $a$
approaches the constant $C=(\gamma_2-\gc)/\gcp$, whose
contours for elliptical core waveguides are shown in Fig.\ \ref{fig1}(iii);
most such wave\-guides have positive $C$ values ranging
up to $400$, but some, those
in the region on the left side of the white curve in Fig.\
\ref{fig1}(iii), have negative or small values of $C$.
The contour plot for $\Delta\beta$ in Fig.\ \ref{fig1}(iv) shows 
that $\Delta\beta$ takes a 
wide range of positive and negative values as $x,y$ vary.

\subsection{Steady state solutions\label{sss1}}

There are four 
classes of steady state solutions of Eqs.\ (\ref{e4},\ref{e5}), each of which
exist only for values of $a,b$ within certain limits, as follows:
\beq
\label{a}
\cos\theta=1, \quad v=\frac{a-1}{2(b-1)}
\eeq
provided $b\ne1$ and $0< \frac{a-1}{2(b-1)}<1$;

\beq
\label{b}
\cos\theta=-1,\quad v=\frac{a+1}{2(b+1)}
\eeq
provided $b\ne-1$ and $0< \frac{a+1}{2(b+1)}<1$;

\beq
\label{c}
\cos\theta=a,\quad v=0
\eeq
provided $|a|\leqslant1$; and

\beq
\label{d}
\cos\theta=-a+2b, \quad v=1
\eeq
provided $|a-2b|\leqslant1$.

\begin{figure}[htbp]
\centering
\includegraphics[width=0.49\columnwidth]{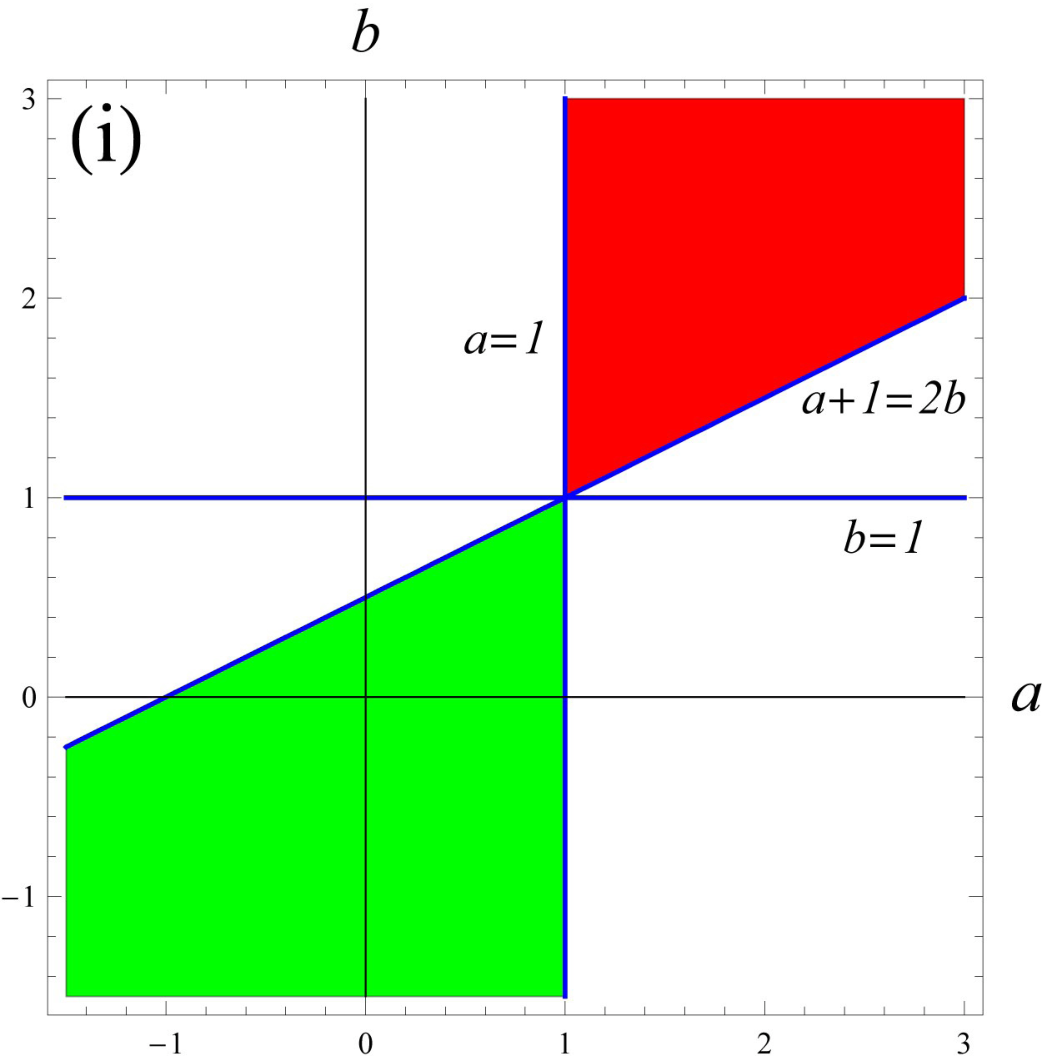}\hfill
\includegraphics[width=0.49\columnwidth]{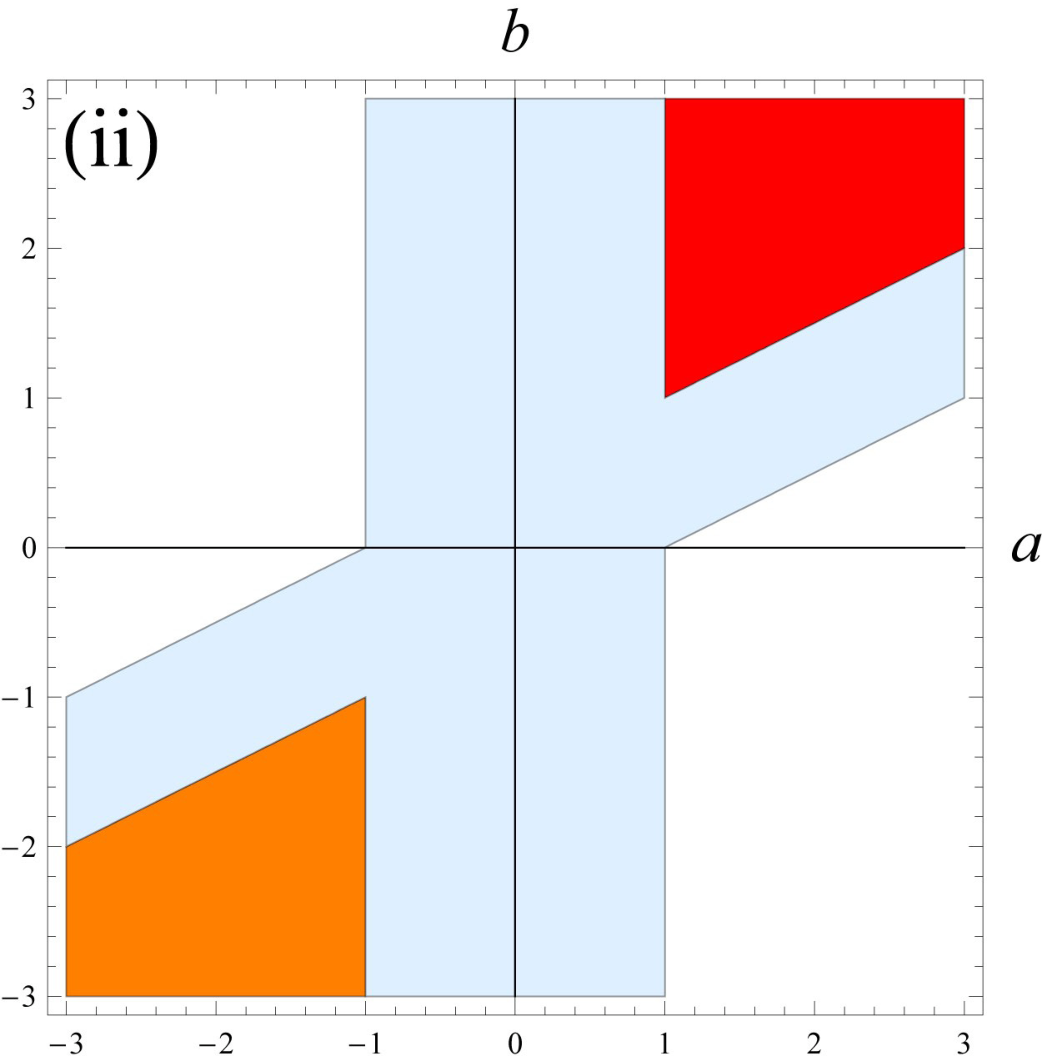}
\caption{The $a,b$ plane showing: (i) the regions of existence for   
the solutions (\ref{a}), either $1<a<2b-1$ (red), or $2b-1<a<1$ (green); 
(ii)  
the regions of existence for the \textit{unstable} 
solutions consisting of (\ref{a}) (red), and (\ref{b}) for
which $2b+1<a<-1$ (orange), 
together with (\ref{c},\ref{d}) for which $|a|<1$ or $|a-2b|<1$ (light blue). 
}
\label{fig2}
\end{figure}

Of these four classes, (\ref{c}) and (\ref{d}) 
lie on the boundary of the physical region
$0<v<1$, but nevertheless influence properties of nearby nontrivial
trajectories, and also play a role in soliton solutions. 
The states (\ref{a}) lie within the physical region only if the 
parameters $(a,b)$
belong to either the red or green region of the $a,b$
plane shown in Fig.\ \ref{fig2} (i). Similarly the solutions (\ref{b}) 
satisfy $0<v<1$ only in the disjoint regions of the $a,b$ plane
defined by either $2b+1<a<-1$ or $-1<a<2b+1$.
If $a,b$ lie outside
these regions, and also outside the strips given by $|a|\leqslant1$ and
$|a-2b|\leqslant1$, there are no steady state solutions.

For special values of $a,b$ these steady states can coincide, for example
if $a=1$ the solution (\ref{c}) coincides with the boundary value of (\ref{a}).
Steady states for values of $a,b$ on the boundary of the regions shown
in Fig.\ \ref{fig2} may
need to be considered separately; for example if $a=b=1$ then all steady states
are given either by (\ref{b}), or else by $\cos\theta=1$ and any constant $v$. 

In practice, the values of $a,b$ are 
determined by the waveguide structure, the propagating mode and, in the case of
$a$, the input power $P_0$, and hence only restricted regions of the 
$a,b$ plane are generally accessible. 
For example, Fig.\ \ref{fig1}(i) shows that for the fundamental mode of 
elliptical core fibers we have $\log_{10}b\geqslant0$,  and so
the attainable values of $b$ are limited to $b\geqslant1$.  
We nevertheless include the case $b<1$ in our analysis, since
this possibility cannot be excluded for other fiber geometries.
We discuss the accessible regions for the case of unstable steady states
in Section \ref{stab}.

\subsection{Lagrangian formulation\label{sss3}}

We wish to determine the stability properties of each of the 
four classes of steady state solutions, in particular we look
for unstable steady states. These are of interest because polarization
states which lie close to these unstable states are very sensitive to
small changes in parameters such as the total power $P_0$, and so can flip
abruptly as a function of the optical fiber length $z$. 
Although we may determine stability properties 
by investigating perturbations about the constant solutions, we find it
convenient to reformulate the defining equations (\ref{e4},\ref{e5}) as the 
Euler-Lagrange equations of a Lagrangian $L$ which is a function of
$\theta,\dot\theta$, and depends otherwise only on the parameters $a,b$.
This also provides insight into the properties and solutions of these equations,
and we may then investigate stability by examining the
corresponding potential function. From (\ref{e5}) we have 
\beq
\label{e10}
v=\frac{\dot\theta+a-\cos\theta}{2(b-\cos\theta)},
\eeq
and by substitution into (\ref{e4}) we obtain
\beq
\label{e11}
2(b-\cos\theta)\;\ddot\theta -\sin\theta\;\dot\theta^2
+\sin\theta(a-\cos\theta)(a-2b+\cos\theta)
=0.
\eeq

We consider 
Lagrangians $L$ of the form
\beq
\label{e14}
L=T-V=\frac{1}{2}M(\theta)\,\dot\theta^2-V(\theta)
\eeq
where $T$ is the (positive) kinetic energy, $V$ is the potential energy, and 
the ``mass" $M$ is a positive function of $\theta$. The equation of motion 
is
\beq
\label{e12}
M(\theta)\,\ddot\theta+\frac{1}{2}M'(\theta)\dot\theta^2+V'(\theta)=0,
\eeq
and is identical to Eq.\ (\ref{e11}) provided
\beq
\label{e13}
M(\theta)=\frac{2}{|b-\cos\theta|},
\qquad
V(\theta)=
-|b-\cos\theta|-\frac{(a-b)^2}{|b-\cos\theta|}.
\eeq

We may therefore investigate all possible solutions $\theta(\tau)$ by
analyzing properties of the periodic potential
$V(\theta)$; 
every solution of the system of equations (\ref{e4},\ref{e5})
corresponds to the trajectory $\theta(\tau)$ of a particle of variable
mass $M$ in the potential $V$. Steady state solutions are
zeroes of $V'(\theta)$, and stability is determined by whether
these zeroes are local maximums or minimums of $V$, subject to the constraint
that the associated function $v$ should always satisfy $0<v<1$.
Trajectories which begin near a local minimum, with a 
small initial speed $\dot\theta(0)$, oscillate periodically with a small 
amplitude. On the other hand, trajectories which begin near
an unstable point, i.e.\ near a local maximum of $V$, can display
periodic oscillations of large amplitude with abrupt transitions 
between adjacent local
maximums; we refer to these as switching solutions (previously bistable 
solutions \cite{Zhang:11}) since 
$\cos\Delta\phi=\cos(\theta/2)$ switches periodically between two distinct
values. Soliton trajectories
also occur in which the particle moves between adjacent local maximums of $V$, 
see for example the discussion in \cite{Coleman}, Section 2 and \cite{KA}
for properties of solitons in optical fibers. As mentioned in
Section \ref{ss3}, soliton trajectories also appear as the separatrix 
in phase plane plots.

\begin{figure}[!ht]
\centering
\includegraphics[width=0.49\columnwidth]{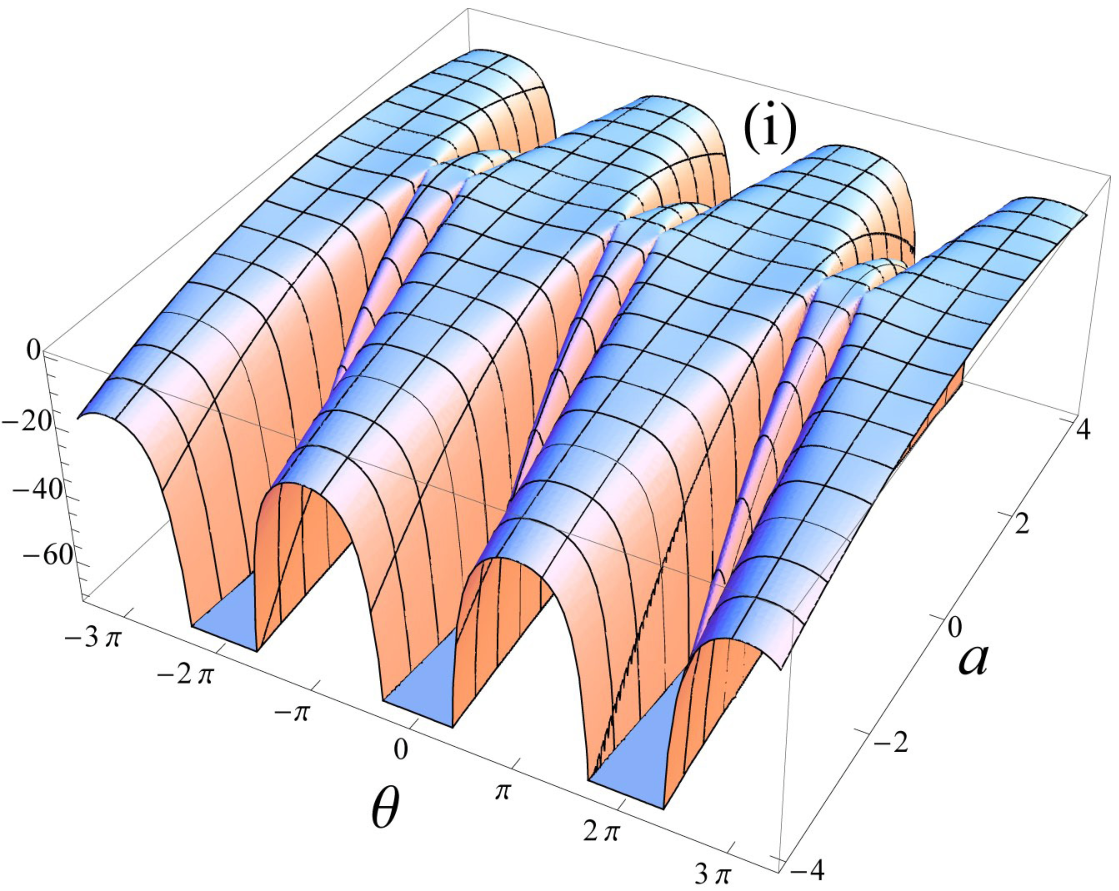}
\includegraphics[width=0.49\columnwidth]{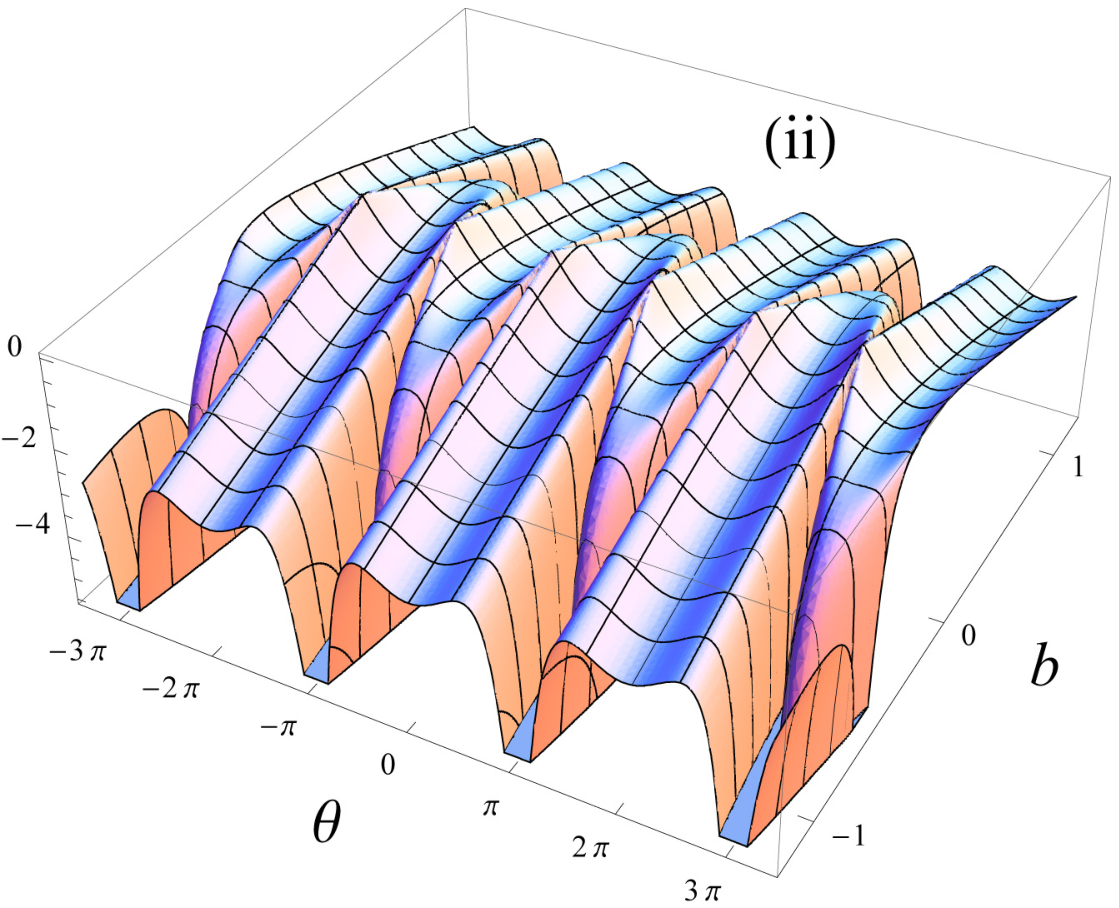}
\caption{The potential $V$ plotted as a function of (i) $\theta,a$ for
$b=0.8$;  (ii) $\theta,b$ for $a=0$.
}
\label{fig3}
\end{figure}

We plot $V$ as a function of $\theta$ and either $a$ or $b$ in Fig.\
\ref{fig3}, showing that $V$ defines
a complex surface with valleys and peaks which change suddenly as $a$ or $b$
are varied. Periodic solutions occur for trajectories restricted to a
local valley, but there are also unbounded trajectories, in which $\theta$
increases or decreases indefinitely, depending on $a,b$ and on
whether $\dot\theta(0)$ is 
sufficiently large. The potential, as a function of $\theta$ and $a$,
has saddle points which indicate that a stable solution can become
unstable as $a$ is varied; according to the definition (\ref{p2}) we may
vary $a$ within certain limits by varying the total power $P_0$.

For $a=b$ the potential is essentially that of the nonlinear pendulum under
the influence of gravity, 
namely a simple cosine potential, but with a mass that depends on
$\theta$. Provided $b>1$ this mass varies between two positive, finite limits. 
The unstable steady states correspond to a pendulum balanced upright,
while the switching states (discussed in Section \ref{ss4}) 
correspond to trajectories which begin with the pendulum positioned 
near the top, possibly with a small initial speed,
then swinging rapidly through $\theta=2\pi$ to reach the adjacent
unstable steady state. During this motion $\cos\Delta\phi=\cos\frac{\theta}{2}$
flips rapidly between the values $\pm1$. The soliton discussed in the
Appendix is the trajectory in which the pendulum begins at the unstable
upright position and, over an infinite time, moves through the stable minimum
to the adjoining unstable steady state.

Although both $M$ and $V$ are singular when $\cos\theta=b$, which occurs
only if $|b|\leqslant1$, this singularity
is an artifact of the Lagrangian formulation, as is evident from 
Eqs.\ (\ref{e4},\ref{e5}), which have smooth bounded right hand sides for any
$b$. In particular $v$, which is obtained from
Eq.\ (\ref{e10}) given $\theta$, is a smooth function of
$\tau$ even if  $\cos\theta=b$ for some $\tau$.

The energy 
$T+V=\frac{1}{2}M(\theta)\,\dot\theta^2+V(\theta)$ is a constant of the
motion.  Hence we may integrate Eq.\ (\ref{e11}) to obtain
\beq
\label{e15}
\dot\theta^2=(b-\cos\theta)^2+(a-b)^2+c\,(b-\cos\theta),
\eeq
where $c$ is the constant of integration. 
This constant is determined by first choosing
initial values $v_0, \theta_0$, where $0<v_0<1$, and then
finding $\dot\theta(0)$ from Eq.\ (\ref{e5}) which,
from (\ref{e15}) evaluated at $\tau=0$, fixes $c$.
We may integrate (\ref{e15}) to determine $\theta$ as an explicit 
function of $\tau$, expressible in terms of elliptic functions, 
as discussed further in Section \ref{ss3}.

A limitation of the Lagrangian formulation is that
the constraint $0<v<1$ is not easily implemented. 
Whereas every solution of the system (\ref{e4},\ref{e5}) defines
a trajectory $\theta(\tau)$ in the Lagrangian system (\ref{e11}), the
converse is not true, i.e\ not all trajectories in this system
satisfy $0<v<1$.
The initial speed $\dot\theta(0)$ must be restricted to only those values
allowed by Eq.\ (\ref{e5}), in which $0<v(0)<1$, and similarly
the constant solutions of Eq.\ (\ref{e11}) are valid
steady states for the original equations (\ref{e4},\ref{e5}) only in certain
regions of the $a,b$ plane.
Trajectories which violate $0<v<1$, while not physical in the context
of optical fiber configurations, can nevertheless be viewed as acceptable
motions of the mechanical system defined by the Lagrangian (\ref{e14}). 
We investigate an alternative Hamiltonian formulation in terms
of $v$ in Section \ref{ss3}.


\subsection{Stability of steady state solutions\label{stab}}

The stability of each of the four classes of steady state solutions
in Section \ref{sss1} is determined
by the sign of $V''$ for that solution; a positive sign implies that the
solution lies at a local minimum of $V$ and is therefore stable, whereas
a negative sign implies that the solution is unstable. 

For the steady states
(\ref{a}) we have $V''=(a-1)(a-2b+1)(b-1)/|b-1|^3$ and so
these states are stable for points $a,b$ such that
$(a-1)(a-2b+1)(b-1>0$, shown as the green region in the $a,b$ plane
in Fig.\ \ref{fig2} (i), and are unstable in the red region, where
$1<a<2b-1$. For the steady states (\ref{b}) we have 
$V''=(a+1)(-a+2b+1)(b+1)/|b+1|^3$ and so
these solutions are stable for $-1<a<2b+1$ and are unstable in
in the orange region $2b+1<a<-1$ shown in Fig.\ \ref{fig2} (ii).  

For the remaining steady states (\ref{c},\ref{d}), 
for which $v=0$ or $v=1$, we have 
$V''=-2\sin^2\theta/|a-b|$ which in all cases 
is negative, and so these states are
unstable whenever they exist. This is consistent with the observation
that $v(\tau)$ cannot attain the values $0,1$ at
any time $\tau$, provided $0<v_0<1$. The regions in the $a,b$ plane where
the unstable states exist are shown in Fig.\ \ref{fig2} (ii).

\begin{figure}[htbp]
\centering
\includegraphics[width=0.49\columnwidth]{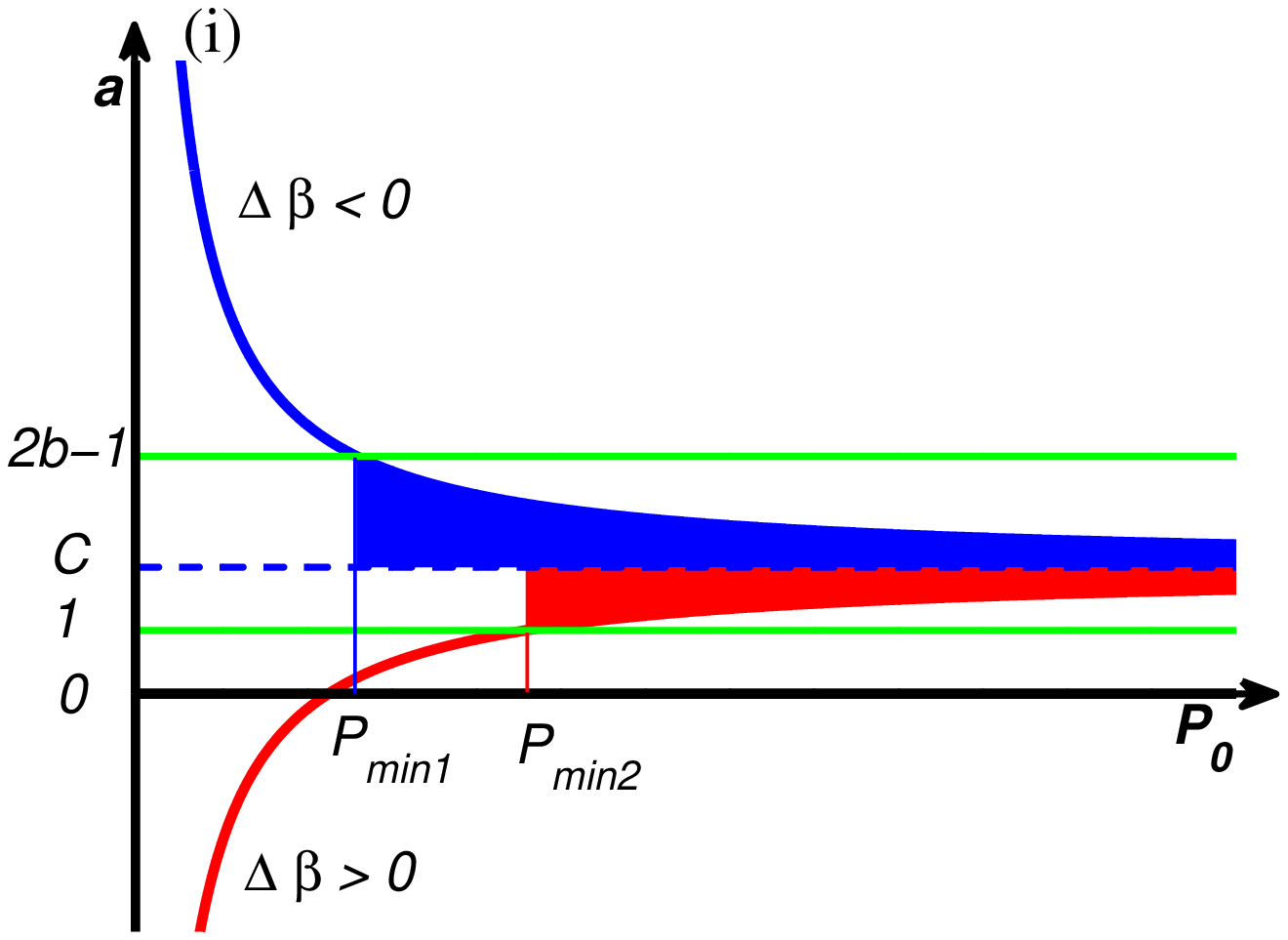}
\includegraphics[width=0.49\columnwidth]{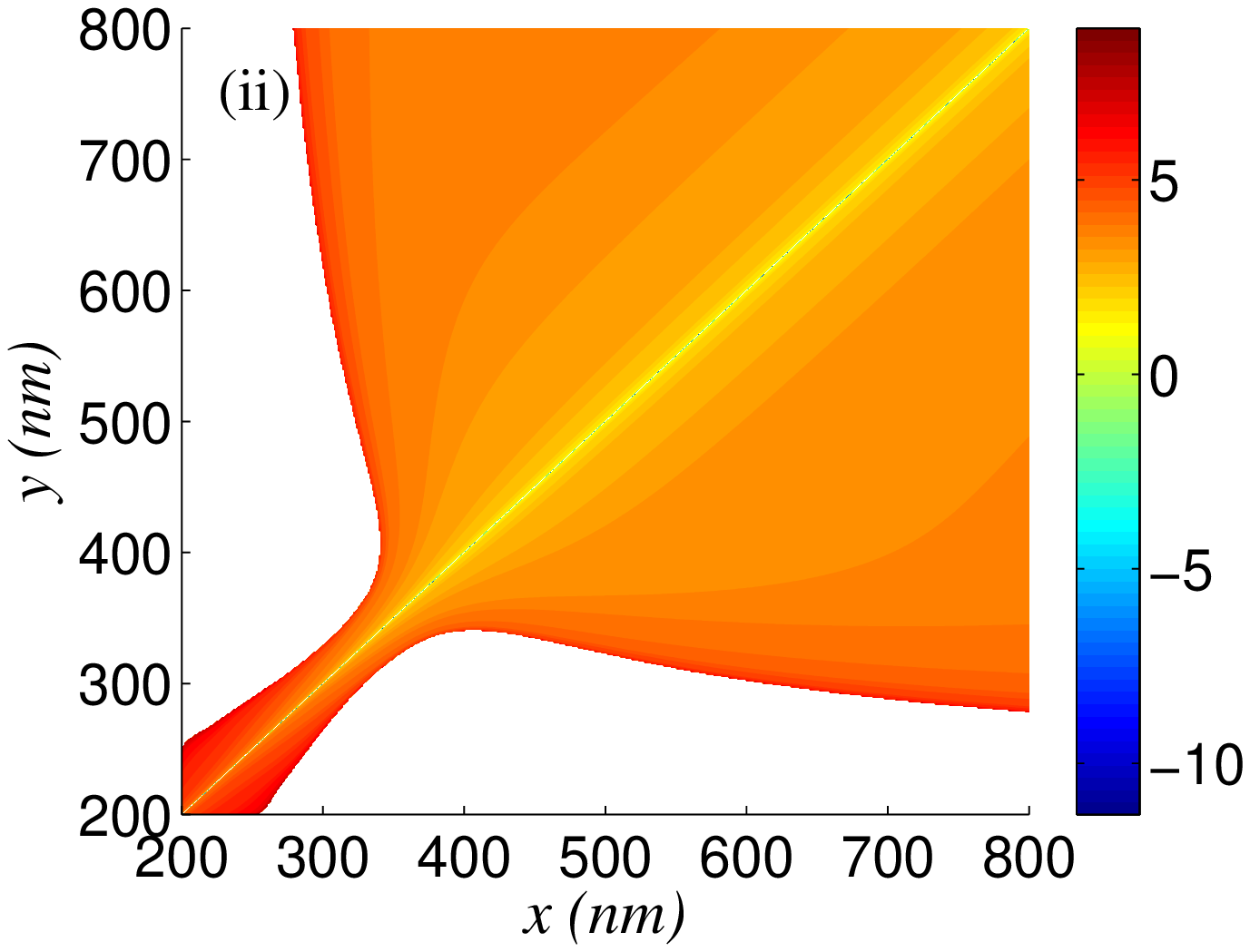}
\caption{
(i) 
$a$ as a function of $P_0$ for $\Delta\beta<0$
(blue solid line) and $\Delta\beta>0$ (red solid line). The green lines
mark the boundaries of the (red) region of instability in the $a,b$ plane shown
in Fig.\ \ref{fig2}(i);
(ii) contour plot of $\log_{10}(P_0^{\min})$ as a function of $x,y$, showing
the minimum total power $P_0^{\min}$ (in units W) required to access unstable 
steady states, where they exist.
}
\label{fig4}
\end{figure}

Next, we determine conditions under which the unstable steady state
solutions (\ref{a}) are accessible.
For elliptical core step index fibers, for which $b>1$ as shown
in  Fig.\ \ref{fig1}(i), the region of instability 
is indeed accessible and leads
to properties such as nonlinear self-polarization flipping, discussed in
Section \ref{ss4}. The region of
unstable solutions is given by $1<a<2b-1$, equivalently
\beq
\label{e36}
\gc+\gcp-\gamma_1<\frac{\Delta\beta}{P_0}<\gamma_2-\gc-\gcp.
\eeq
These inequalities specify the possible values, if any, of $P_0$ for which 
the unstable solutions exist for a fixed fiber. 
In order to visualize this region we
plot $a$ as a function of $P_{0}$ in Fig.\ \ref{fig4}(i), where $a$ 
is given by (\ref{p2}). 
The boundaries of the unstable region at $a=1,a=2b-1$ are shown by the
green solid lines. 

First we consider fibers for which $1<C<2b-1$, where
$C=(\gamma_2-\gc)/\gcp$ takes the value shown by the dashed line
in Fig.\ \ref{fig4}(i). Then $a$ has two
branches associated with either $\Delta\beta<0$ or $\Delta\beta>0$;
for the branch corresponding to $\Delta\beta<0$ (the solid blue line), 
$a$ is large and positive for small $P_0$ and 
asymptotically approaches $C$ for large $P_0$. 
The intersection of this branch with the boundary $a=2b-1$
determines the minimum power $P_{\min_1}$ required in order to
access the unstable region. In this case, only part of the unstable region
corresponding to $C<a<2b-1$ is accessible, as shown by the blue region. 
For the 
$\Delta\beta>0$ branch (red solid curve) $a$ is large and negative for 
small $P_0$
and asymptotically approaches $C$ for large $P_0$. For this branch, $P_{0}$ 
needs to be larger than a value $P_{\min_2}$. The unstable region is accessible
provided $1<a<C$ and is a subset (red shaded) of the whole
unstable solution region. Fig.\ \ref{fig4}(i) allows one to
determine the minimum and maximum values of $a$ and the minimum power to
access the unstable solution region, once $\Delta\beta$ and $C$ are
known. For elliptical core fibers these two values are completely determined 
by  the dimensions $x,y$, see Fig.\ \ref{fig1}(iii,iv) for plots of $C$ and
$\Delta\beta$. 

Besides fibers for which $1<C<2b-1$, there are the
possibilities $C>2b-1$ or $C<1$. From Fig.\ \ref{fig1}(iii,iv) one can 
show that these combinations (with $\Delta\beta$ positive or 
negative) either do not exist, or do not lead to unstable solutions,
since the possible values of $a$ do not
lie in the unstable region $1<a<2b-1.$ In summary, the only
elliptical core fibers that allow unstable solutions are those with $1<C<2b-1$ 
with either positive or negative $\Delta\beta$. The case in which
$\Delta\beta=0$ is discussed separately in \cite{ZLMS, ZLMS2}.

Based on the above discussion, one can find the minimum power $P_0^{\min}$
required to generate unstable solutions for elliptical core fibers.
Fig.\ \ref{fig4}(ii) plots 
$\log_{10}(P_0^{\min})$ (where $P_0^{\min}$ is measured in watts)
as a function of $x,y$, where the white region 
corresponds to fibers for which there are no unstable 
solutions, and the regions below and above the diagonal line 
correspond to $P_{\min_1}$ and $P_{\min_2}$, respectively, which 
have been obtained for the two branches of the function $a(P_{0})$
shown in Fig.\ \ref{fig4}(i).

\subsection{\label{ss3}Hamiltonian function}

Although the Lagrangian formulation in terms of $\theta$ is convenient for an
analysis of the steady states and their stability, and also for a qualitative 
understanding of all solutions including solitons, 
the constraint $0<v<1$ is more easily implemented by means of a 
direct formulation in terms of $v$. This automatically eliminates unphysical 
trajectories for which one of the input powers $P_1,P_2$ is negative.
Such a formulation follows by construction of a Hamiltonian function which,
being conserved, allows us to firstly integrate the nonlinear equations 
and obtain analytical solutions and, secondly, to interpret physically
the possible states of polarizations within an optical waveguide 
from the phase plane contours. Corresponding to the conserved energy $T+V$ 
which follows from (\ref{e14})
there is a Hamiltonian function $H$ defined by
\beq
\label{e33}
H(v,\theta)=-av+bv^2+v(1-v)\cos\theta
\eeq
which satisfies
\[
\dot v=-\frac{\partial H}{\partial\theta},
\qquad
\dot\theta=-\frac{\partial H}{\partial v}.
\]
Hence as a function of $\tau$, $H$ is conserved  and takes the constant
value $H_0=H(v_0,\theta_0)$ on any trajectory. We may investigate
all possible solutions, therefore, by analyzing the curves of constant
$H_0$ in the $v,\theta$ plane. We have
\beq
\label{e16}
\cos\theta=\frac{H_0+av-bv^2}{v(1-v)},
\eeq
and from Eq.\ (\ref{e4}) we obtain
\beq
\label{e17}
\dot v^2=Q(v),
\eeq
where $Q$ is the polynomial of 4th degree (provided $b^2\ne1$) given by
\beq
\label{e34}
Q(v)=v^2(1-v)^2-(H_0+av-bv^2)^2.
\eeq 
Since the left hand side of (\ref{e17}) is positive, solutions
exist only if $Q(v)\geqslant0$ for $v$ in the interval $0<v<1$. 
Generally $Q(0),Q(1)<0$ but since
$Q(v_0)=v_0^2(1-v_0)^2\sin^2\theta_0\geqslant0$
(as follows from Eq.\ (\ref{e4})) $Q$ has at least two real zeroes,
possibly repeated, and so there is an interval within $0<v<1$ in which
$Q(v)>0$, and so solutions always exist.
If the initial values $v_0,\theta_0$ are such that the trajectory
begins in a stable steady state, $v$ remains constant for all $\tau>0$,
otherwise the trajectory is nontrivial. There are two types of
nontrivial solutions, periodic and soliton solutions.

We can gain insight into possible solutions by plotting contours of 
constant $H(v,\theta)$ in the $v,\theta$ plane, which supplies
essentially a phase portrait
of the system. Solutions for which both $v,\theta$ are periodic in $\tau$
form closed loops, and lie close to a stable steady state, whereas nonperiodic
trajectories lie outside the separatrix which defines soliton
solutions, as we discuss in the Appendix. Fig.\ \ref{fig5} shows
two examples in which stable steady states are marked in green, and
unstable steady states are shown in red or orange. Periodic solutions
are evident as closed loops surrounding stable steady states, whereas the
separatrix marks soliton trajectories which connect unstable
steady states. Apart from these solitons, all other solutions $v,\cos\theta$
(but not necessarily $\theta$) are periodic in $\tau$. The switching
solutions of particular interest, in which the state of polarization 
inside the waveguide flips between two well-defined states, 
are those close to the separatrix.

\begin{figure}[htbp]
\centering
\includegraphics[width=0.49\columnwidth]{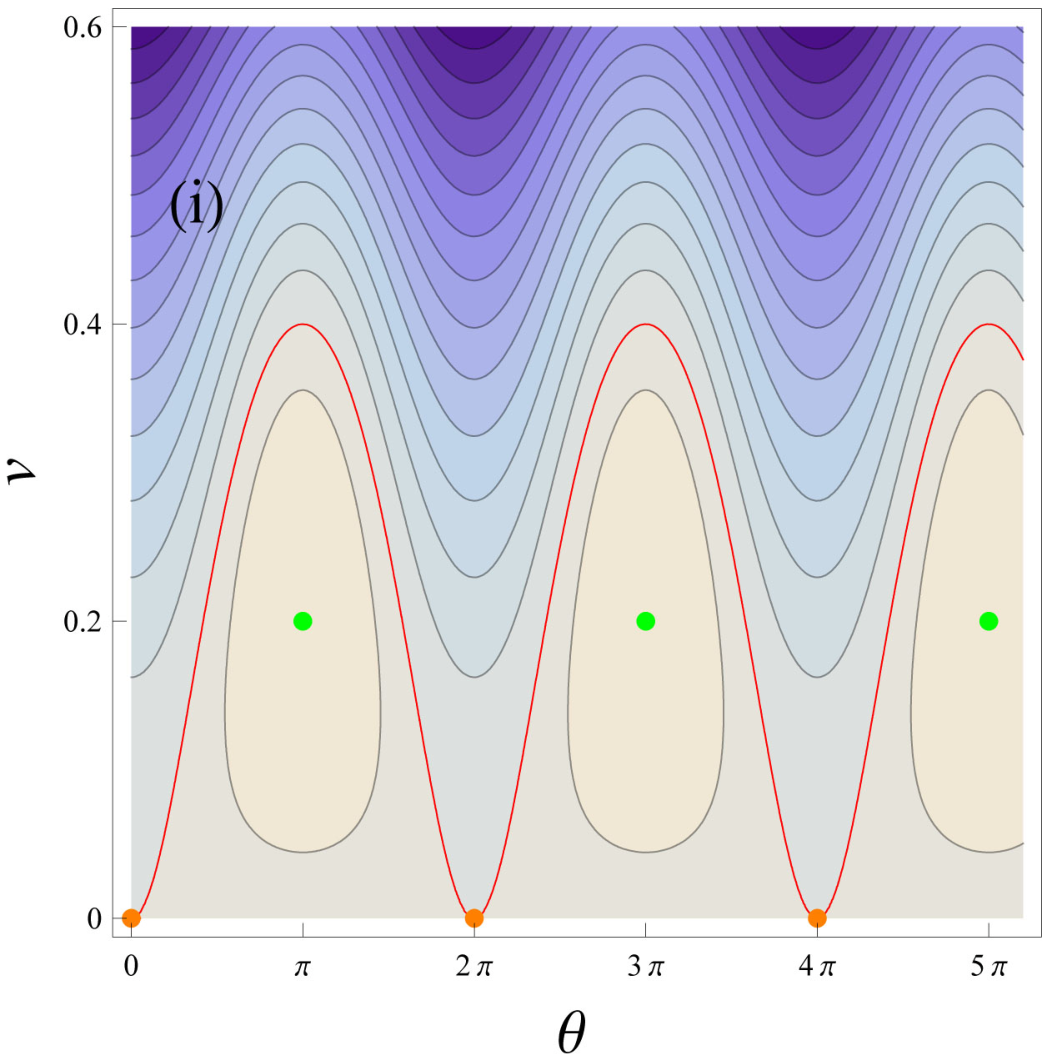}\hfill
\includegraphics[width=0.49\columnwidth]{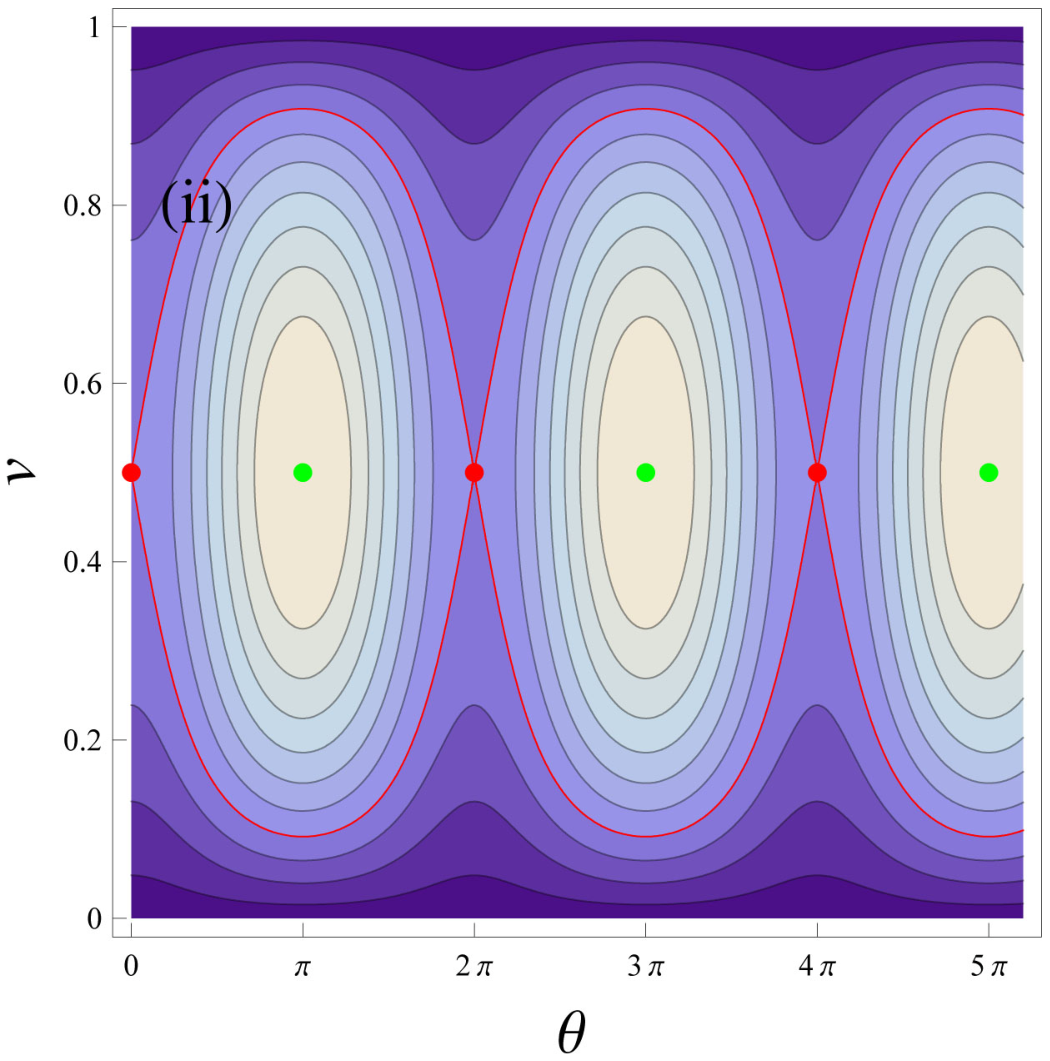}
\caption{Contours in the $\theta,v$ plane of constant $H$ for
(i) $a=1,b=4$; (ii) $a=b=2$, with steady states marked by green dots (stable)
and red or orange dots (unstable). The separatrix, which identifies the
soliton trajectories, is shown in red.
}
\label{fig5}
\end{figure}


\section{\label{ss4}Periodic solutions}

Periodic solutions  $v$ of (\ref{e17})
attain both minimum and maximum values, denoted $\vmin, \vmax$
respectively, with $0<\vmin\leqslant\vmax<1$. Since $\dot v=0$ at
a maximum or minimum of $v$, both $\vmin, \vmax$ are roots of $Q$.
We can factorize $Q$ as a product of quadratic polynomials,
\beq
Q(v)=-\left[(b+1)v^2-(a+1)v-H_0\right]
\left[(b-1)v^2-(a-1)v-H_0\right],
\eeq
and hence explicitly find all roots, 
and so identify $\vmax$ and $\vmin$.
We integrate
${\dot{v}}=\sqrt{Q(v)}$ over the half-period in which $v$ increases, in order
to find $\tau$ as a function of $v$, and also the period $T$:
\begin{equation}
\label{e19}
\int_{v_{\mathrm{min}}}^{v}\frac{du}{\sqrt{Q(u)}}
=\tau-\tau
_{0},\quad T=2\int_{v_{\mathrm{min}}}^{v_{\mathrm{max}}}
\frac{du}{\sqrt{Q(u)}},
\end{equation}
where $\tau_0$ is the time at which  $v$ achieves its minimum, i.e.\ 
$v_{\mathrm{min}}=v(\tau_{0})$. These integrals may be evaluated in
terms of elliptic integrals of the first kind, see for example the explicit
formulas in \cite{GR}
(Sections 3.145, 3.147). In particular, $T$ is expressible in terms of the
complete elliptic integral $K$, and so can be written as an explicit function
of $a,b,v_{0},\theta_{0}$, i.e.\ as a function of the wave\-guide parameters
and the initial power and phase of the input fields. The precise formulas
depend on the relative location of the roots of $Q$.

Having found $v$, $\cos\theta$ is obtained from Eq.\ (\ref{e16}) and
is also periodic in $\tau$, as is $\dot\theta$ which is obtained
from Eq.\ (\ref{e5}), however $\theta$ itself need not 
be periodic.
Although it is straightforward to find $v,\theta$ numerically as functions
of $\tau$, for specified numerical values of $a,b$ and initial values
$v_0,\theta_0$, the exact solutions are useful because they display the exact
dependence of the solution on all parameters, such as the total power $P_0$;
it is not necessary therefore
to solve the equations numerically for every choice of $P_0$,
rather the exact solution  gives the explicit periodic solution
and the period as known functions of $P_0$.  

For switching solutions, the phase difference between the two polarization
vectors experiences abrupt phase shifts through $\pi$ as the light 
propagates within the waveguide. As a result, the 
state of polarization flips between two well-defined polarization 
states, where the flipping angle depends on $a,b$ and on $\theta_0,v_0$. 
The following are two examples of switching solutions.

\begin{figure}[!ht]
\centering
\includegraphics[width=0.49\columnwidth]{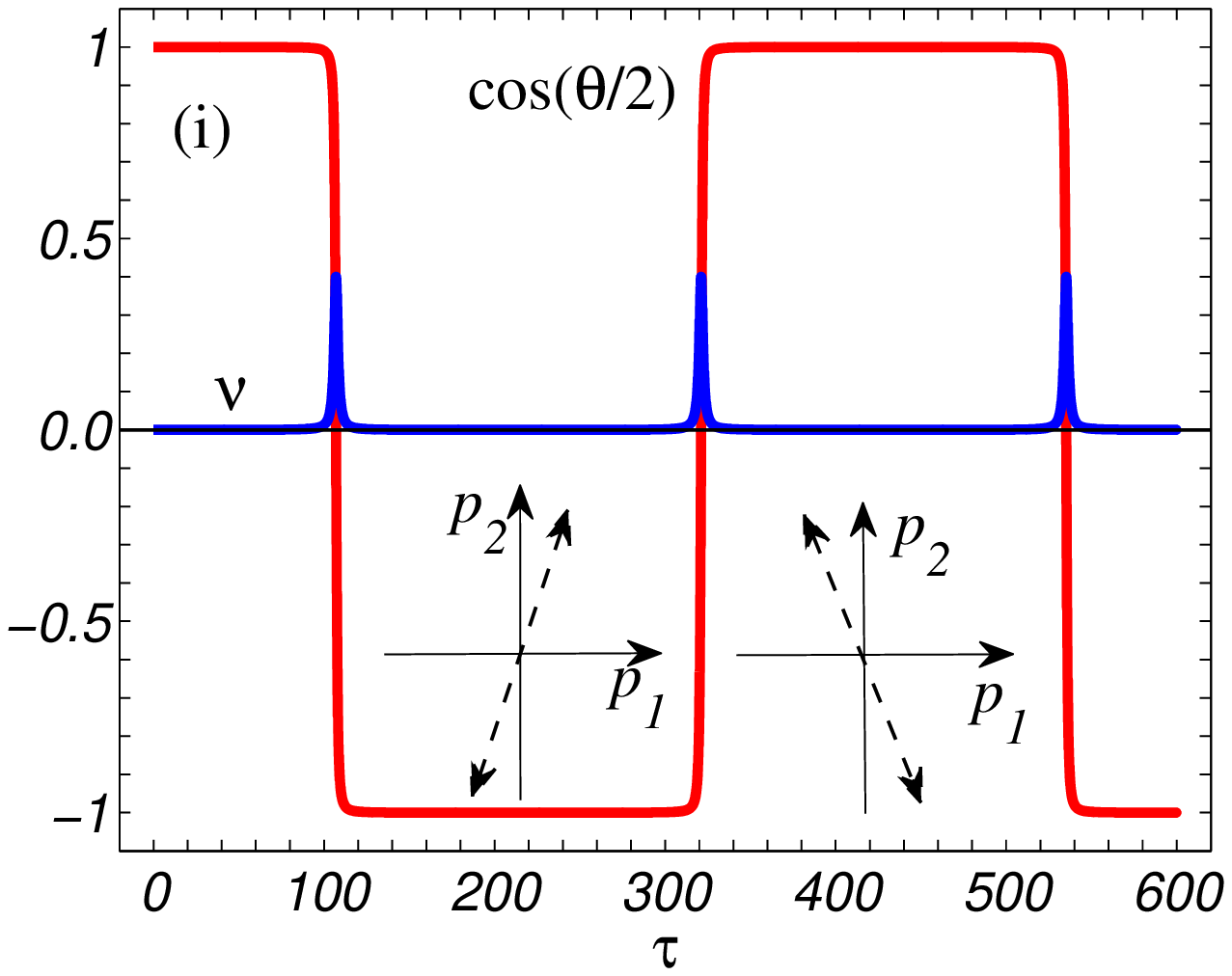}
\includegraphics[width=0.49\columnwidth]{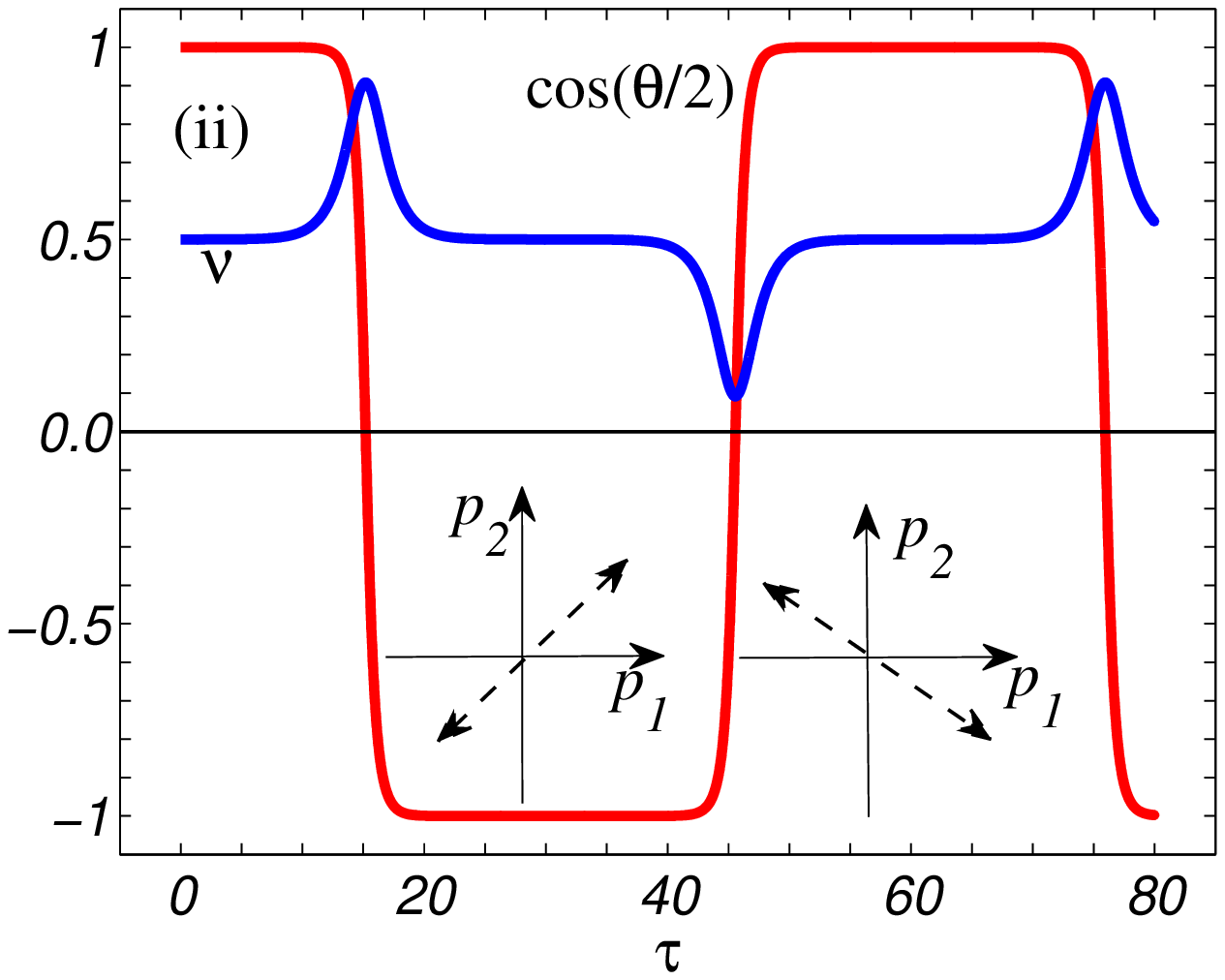}
\caption{
Switching solutions $v$ and $\cos\frac{\theta}{2}=\cos\Delta\phi$ as 
functions of $\tau$
for: (i) $a=1,b=4$ and $v_0=\varepsilon,\theta_0=0$;
(ii) $a=b=2$ and
$v_0=\frac{1}{2}, \theta_0=\varepsilon$ where $\varepsilon=10^{-4}$.
The insets show the polarization vectors associated with 
the values $\cos\Delta\phi=\pm1$.
}
\label{fig6}
\end{figure}

As the first example we choose $a=1, b=4$ with the initial
values $v_0=\varepsilon, \theta_0=0$, where $\varepsilon=10^{-4}$, in which
case
the input laser beam is linearly polarized and the polarization state is 
close to one of the principle axes of the waveguide.  
Hence, the trajectory starts near the unstable steady states (\ref{a})
or (\ref{c}), which lie on the boundary of the red region shown in
Fig.\ \ref{fig2} (i).
We plot $v$ and $\cos\frac{\theta}{2}=\cos\Delta\phi$ 
as a function of $\tau$ in
Fig.\ \ref{fig6} (i), showing switching behavior for $\cos\frac{\theta}{2}$, 
which is periodic and flips abruptly between the values $\pm1$;
$\theta$, however, is an increasing function of $\tau$, with jumps through
$2\pi$ at periodic intervals. The polarization vector 
experiences an angular flipping associated with the abrupt flipping of 
$\cos\Delta\phi$, however, since $v_0=\epsilon$ and 
$\theta_0=0$, the flipping angle is very small,
as depicted in the inset of Fig.\ \ref{fig6} (i).
Regarded as the trajectory of a particle of
mass $M$ in the potential $V$ in Eq.\ (\ref{e13}) this
motion corresponds to a particle moving slowly over the peaks
of the potential, which are the unstable steady states, 
then sliding quickly down the valleys
through the minimum values of $V$ and back to the
peaks. For $a=1$ the potential is flat at
its maximum values, since in this case $V'=0=V''=V'''$, 
hence $v, \dot\theta$ are
each close to zero except when $\theta$ moves to an adjoining maximum of $V$.
In terms of the contour plots shown in Fig.\ \ref{fig5}(i) this trajectory
corresponds to the contour which begins just above the unstable
steady state (orange dot) and closely follows the separatrix shown in red
(which is the
soliton solution discussed in the Appendix) with a maximum value
$\sim0.4$ for $v$.

As a second example of switching behavior we choose $a=b=2$ with
$v_0=1/2, \theta_0=\varepsilon$, where $\varepsilon=10^{-4}$, which
corresponds to a linearly polarized input laser beam in which the polarization 
vector makes an angle of $45^{\circ}$ to either of the principle axes of the 
waveguide. Again, the initial value lies close
to an unstable steady state (\ref{a}) and $a,b$ lie within the
red region of instability in Fig.\ \ref{fig2}.
We plot $v$ and $\cos(\theta/2)$ as functions of $\tau$ in Fig.\ 
\ref{fig6} (ii), showing the periodicity of these functions
and the switching behavior of $\cos(\theta/2)$. 
Since $v_0=1/2$, the angular flipping of the polarization vector is $\pi/2$,
because 
$\cos(\theta/2)$ flips between values $\pm1$, as shown in the inset of 
Fig.\ \ref{fig6} (ii).
Unlike the previous example, $\theta$ is also periodic in $\tau$ with a
trajectory that corresponds to the motion of a particle in the potential
$V$, starting
slowly near the unstable steady state (\ref{a}) but sliding rapidly through
the potential minimum to approach an adjoining unstable steady state. This
motion is similar  to the periodic oscillations of a nonlinear pendulum 
(since $a=b$, see the definition of $V$ in Eq.\ (\ref{e13}))
with a large amplitude of almost $2\pi$, and
$v$ attains nearly all values between $0,1$.
In terms of the phase space contours shown in Fig.\ \ref{fig5}(ii),
the motion corresponds to a periodic trajectory which begins near the red
dot (unstable steady state) and again closely follows the separatrix which
marks the soliton trajectory.


\section{\label{s9}Discussion and Conclusion}

Switching states, as defined and demonstrated here through simulation by means
of a full vectorial model, are attractive for practical applications,
since they allow nonlinear self-flipping of the polarization
states of light propagating in an optical waveguide. This flipping is due to
the nonlinear interactions of the two polarizations, and has properties that 
depend on the total optical power and on the specific fiber parameters. 
These properties can in principle be employed to 
construct devices such as optical logic gates \cite{Zhang2010}, 
fast optical switches and optical limiters \cite{ZLMS,ZLMS2}, in which 
small controlled changes in the input parameters lead to sudden changes in 
the polarization states. 

The minimum power necessary to
generate such switching states is determined for any waveguide by 
the inequalities (\ref{e36}) and, for
chalcogenide optical nanowires with elliptical core cross sections, is 
summarized in Fig. \ref{fig4} (ii). The minimum
power required in such nanowires is in the range $1-10$kW which, although
not practicable for CW lasers, can be achieved in pulsed lasers.
Although we have limited our analysis to the static case,
ignoring the temporal variation of laser light, it is still applicable to
slow pulses with pulse widths in the order of nanoseconds depending
on the dispersion of the waveguide. A more practical minimum power requirement
that achieves switching behavior is by means of asymmetric waveguides, 
such as rib waveguides, for which $\Delta\beta$ can be
reduced to very small values
while still having different field distributions for the two
polarizations, as discussed in \cite{ZLMS,ZLMS2}. 

The nonlinear interactions of the two polarizations can be impacted by
two factors that have not yet been investigated: 
(1) interactions with higher order
modes in few-moded waveguides and, (2)  contributions from nonlinear
terms containing different forms of $\mathbf{e}_1\centerdot\mathbf{e}_{2}$,
i.e., nonzero values for the coefficients 
$\gamma_{\mu\nu}^{(1)},\gamma_{\mu\nu}^{(2)},\gamma_{\nu}^{(3)}$ 
in Eqs.\ (\ref{100c}-\ref{100e}). (This applies only when  
$\mathbf{e}_1\centerdot\mathbf{e}_{2}$ is no longer approximately zero,
as assumed in this paper).
In few-moded waveguides, higher order modes contribute to
the nonlinear phase of each polarization of the fundamental mode through cross
phase and coherent mixing terms. Inspection of Eq.\ (\ref{01}) reveals that
nonzero 
$\gamma_{\mu\nu}^{(1)},\gamma_{\mu\nu}^{(2)},\gamma_{\mu\nu}^{(3)}$ 
coefficients significantly change the dynamics of nonlinear
interactions of the two polarizations and most likely lead to different 
parameter regimes
for the existence of periodic and solitonic solutions. 
These factors will be the subject of further studies.

In summary, we have developed the theory of nonlinear interactions of the
two polarizations using a full vectorial model of pulse propagation in high
index subwavelength waveguides. This theory indicates that there is an
anisotropy in the nonlinear interactions of the two polarizations that
originates solely from the waveguide structure. 
We have found all static solutions of the nonlinear system of
equations by finding exact constants of integration, which leads to
expressions for the general solution in terms of elliptic functions.
We have analyzed the stability of the
steady state solutions by means of a Lagrangian
formalism, and have shown that there exist periodic switching solutions, 
related to a class of unstable steady states, for which
there is an abrupt flipping of the polarization states through an angle 
determined by the
structural parameters of the waveguide and the parameters of the input laser.
By means of a Hamiltonian formalism we have analyzed all solutions,
including solitons which we have shown are close to the switching solutions 
of interest.

\section*{Acknowledgements}

This research was supported under the Australian Research Council's Discovery
Project funding scheme (project number DP110104247). Tanya M. Monro
acknowledges the support of an ARC Federation Fellowship.


\section*{Appendix}

We include here a discussion of the topological
solitons which appear as solutions of Eq.\ (\ref{e11}), as 
configurations $\theta(\tau)$ which interpolate between the adjacent maximums
of the periodic potential $V$ defined in Eq.\ (\ref{e13}). 
They define trajectories which
move between adjacent unstable steady states with abrupt transitions, to form
``kinks" which are stable against time-dependent perturbations.
Such trajectories are visible in Fig.\ \ref{fig5} (i) and (ii)  (the 
contours marked in red)
as they form the separatrix between periodic solutions $v,\theta$ and
nonperiodic solutions. The fact that solitons can occur in this way has been
previously noted, see for example Chapter 9 in \cite{KA}.
In Fig.\ \ref{fig5} (i) the soliton is the trajectory which connects the
adjacent unstable steady states (orange) at $v=0$ and 
$\theta=0, 2\pi, 4\pi \dots$ and
similarly in Fig.\ \ref{fig5} (ii) the solitons connect the
(red) unstable steady states.
Such solutions exist on the full real line $-\infty<\tau<\infty$, with 
appropriate boundary conditions, but are also solutions
on any finite subset of the real line, 
corresponding to an optical fiber of finite length, 
with boundary values obtained from the exact solution.

Solitons are significant in the context of switching solutions since switching
behavior occurs precisely when solutions lie near soliton trajectories;
the switching solutions shown in Fig.\ \ref{fig6} (i) and (ii), for example, 
correspond to contours in Fig.\ \ref{fig5} (i) and (ii) which lie very close to
the separatrix. The soliton itself is 
not periodic but nearby trajectories are periodic for both $v$ and 
$\cos\theta$ as functions of $\tau$. The abrupt transitions which characterize
switching, as shown for example in Fig.\ \ref{fig6}, can equally be viewed as
the ``kinks" of a soliton, in which $\cos(\theta/2)$ changes between two
distinct values over a very short $\tau$-interval, and in doing so 
interpolates between unstable steady states.
We are interested here mainly in transitions between the unstable steady states
(\ref{a}), since these correspond to polarization flipping, i.e.\ 
$\cos\Delta\phi=\cos(\theta/2)$ flips between values $\pm1$. There exist,
however, solitons corresponding to the other unstable steady states such
as (\ref{c},\ref{d}), which we also discuss briefly.

In order to find explicit solutions, we define a 
potential $U$ according to $U(\theta)=V_0-V(\theta)$, 
where the shift $V_0$ is selected such that
the minimum value of $U$ is zero.  If $1<a<2b-1$, for example, in which
case the unstable steady states (\ref{a}) exist, we have
\beq
\label{e31}
V_0=1-b-\frac{(a-b)^2}{b-1}.
\eeq
We also define the positive ``action" functional $S$ by
\beq
\label{e26}
S(\theta,\dot\theta)=\int_{-\infty}^{\infty}
\left[
\frac{1}{2}M(\theta)\;\dot\theta^2+U(\theta)
\right]d\tau.
\eeq
Eqs.\ (\ref{e11}) and (\ref{e12}) follow by using Hamilton's principle of
least action applied to $S$. We can write
\beq
S=\int_{-\infty}^{\infty} 
\frac{1}{2}M\left[\dot\theta\mp\sqrt{\frac{2U}{M}}\right]^2 \,d\tau
\pm
\int_{-\infty}^{\infty} 
M\sqrt{\frac{2U}{M}}\;\dot\theta\,d\tau.
\eeq
The last term takes values only on the boundary and so does not vary as
$\theta,\dot\theta$ are varied, hence a local minimum of $S$ occurs when
\beq
\label{e27}
\dot\theta=\pm\sqrt{\frac{2U}{M}},
\eeq
which implies $M\dot\theta^2=2U$. Solutions of this equation, which is
equivalent to Eq.\ (\ref{e15}) with $c=V_0$, satisfy 
Eqs.\ (\ref{e11}) and (\ref{e12}) with the property that $S<\infty$.
Hence, for such solutions we have $\dot\theta\to0$
and $\theta$ approaches a zero of $U$ as $|\tau|\to\infty$.
We therefore integrate Eq.\ (\ref{e27}) or equivalently
Eq.\ (\ref{e15}) with $c=V_0$.

For the first example we select $a,b$ in the red region in Fig.\
\ref{fig2} for which $1<a<2b-1$, with $c=V_0$ given by (\ref{e31}), 
then the soliton
interpolates between the unstable steady states (\ref{a}).
By direct integration of Eq.\ (\ref{e15}) or (\ref{e27}) we obtain
\beq
\label{e28}
\cos\theta
=
1+\frac{2\kappa}
{1-(\kappa+1)\cosh^2\sqrt{\kappa}\,(\tau-\tau_0)},
\eeq
where $\tau_0$ is the constant of integration, and
\[
\kappa=\frac{(a-1)(-a+2b-1)}{2(b-1)}.
\]
The solution satisfies $\lim_{|\tau|\to\infty}\cos\theta=1$ and at 
$\tau=\tau_0$, which may be regarded as the location of the soliton, 
we have $\cos\theta=-1$. By suitable choice of sign for $\theta$,
and by choice of the branch of the inverse cosine function, 
we obtain $\theta$ as a function of $\tau$
which either increases or decreases between any 
two adjacent zeros of the potential $U$ at $\cos\theta=1$. 
From Eq.\ (\ref{e10}) we obtain $v$:
\beq
\label{e32}
v=\frac{a-1}{2(b-1)}
+\frac{\kappa}
{a-b\pm(b-1)\sqrt{\kappa+1}\,\cosh\sqrt{\kappa}\,(\tau-\tau_0)},
\eeq
where the sign corresponds to either increasing or decreasing $\theta$,  
and we have $\lim_{|\tau|\to\infty}v(\tau)=\frac{a-1}{2(b-1)}$.

As a specific example, for $a=b=2$ and $\kappa=1/2$,
the separatrix trajectory shown in Fig.\ \ref{fig5} (ii) is 
the parametric plot of $v,\theta$ as functions of the parameter $\tau$; 
$v$ evidently varies between maximum and 
minimum values which occur at $\tau=\tau_0$, as
can be determined directly from (\ref{e32}).
We can also find the solutions (\ref{e28},\ref{e32}) directly by solving
Eq.\ (\ref{e17}). It is necessary only to determine
$H_0=H(v_0,\theta_0)$ by choosing $v_0,\theta_0$ at $|\tau|=\infty$, which then
determines $Q$ from (\ref{e34}). For the states
(\ref{a}) we  obtain $H_0=-\frac{(a-1)^2}{4(b-1)}$ and $Q(v)$ has a 
repeated root at $v=\frac{a-1}{2(b-1)}$; the expression (\ref{e32}) for
$v$ may then be obtained by using the general
integration formulas in Sections 2.266, 2.269 of Ref.\ \cite{GR}.

Solitons also exist corresponding to the unstable steady states (\ref{b}),
provided $2b+1<a<-1$ and $b<-1$, and may be obtained
from the formulas (\ref{e28},\ref{e32}) by 
means of the symmetry 
$\tau\to-\tau,\theta\to\theta+\pi,a\to-a,b\to-b$ which leaves Eqs.\
(\ref{e4},\ref{e5}) invariant. 
The parameter $\kappa$, for example, 
is now defined  by $\kappa=(a+1)(a-2b-1)/2/(b+1)$ which
is positive in the orange region of Fig.\ \ref{fig2} (ii). 

Consider next the unstable states (\ref{c}), which are defined only in the 
strip $|a|\leqslant1$ of the $a,b$ plane. 
Soliton solutions take the values
$\cos\theta=a,v=0$ as $|\tau|\to\infty$, and hence the Hamiltonian
function $H(v,\theta)$ defined in Eq.\ (\ref{e33}) takes the constant 
value $H_0=0$, which corresponds to $c=V_0=2(a-b)$ in Eq.\ (\ref{e15}).   
By solving $\dot v^2=Q(v)$ we find:
\beq
\label{e35}
v(\tau)=\frac{1-a^2}
{1-ab+|b-a|\,\cosh[\sqrt{1-a^2}\,(\tau-\tau_0)]},
\eeq
which exists for all $|a|<1$ and $b\ne a$. We have
$\lim_{|\tau|\to\infty}v(\tau)=0$ and
$v$ attains its maximum value $\vmax$
at $\tau=\tau_0$, with either $\vmax =(a+1)/(b+1)$ for $b>a$ or else
$\vmax =(a-1)/(b-1)$ for $b<a$. 
Having found $v$, we obtain $\cos\theta$
from Eq.\ (\ref{e16}) with $H_0=0$ using $\cos\theta=(a-bv)/(1-v)$, 
specifically
\beq
\label{e37}
\cos\theta(\tau)=a-\frac{1-a^2}
{-a +\eta\cosh[\sqrt{1-a^2}\,(\tau-\tau_0)]},
\eeq
where $\eta=(b-a)/|b-a|$ is the sign of $b-a$.
We have $\dot\theta=a-\cos\theta$ and $\cos\theta(\tau_0)=-\eta$. 
For the special case $b=a$ with $|a|<1$, or if $a=1$, we solve 
$\dot v^2=Q(v)$ directly; in the latter case we obtain
\beq
\label{e38}
v(\tau)=\frac{2}{b+1+(b-1)(\tau-\tau_0)^2},
\qquad
\cos\theta=1-\frac{2}{1+(\tau-\tau_0)^2}.
\eeq

As a specific example we choose $a=1, b=4$, for which contour plots for 
constant $H$ are shown in Fig.\ \ref{fig5} (i); the (red)
separatrix trajectory in particular
is visible as the curve which connects the unstable steady states at
$v=0, \theta=0,2\pi \dots$. This separatrix is precisely
the parametric plot of $v,\theta$ given by (\ref{e38}), where
$v$ evidently varies between zero and its
maximum value of $2/(b+1)=0.4$ which occurs at $\tau=\tau_0$, while
$\cos\theta$ varies between the values $1$ as $|\tau|\to\infty$,
 when $v=0$, and $-1$ at $\tau=\tau_0$.

There are also solitons corresponding to the unstable steady states
(\ref{d}). Precise formulas can be obtained from Eqs.\ (\ref{e35},\ref{e37}) 
by means of the transformations $\theta\to-\theta,v\to1-v,a\to-a+2b$
which are discrete symmetries of the defining equations (\ref{e4},\ref{e5}).

\end{document}